\documentclass[]{aa} 

\usepackage{graphicx}	
\usepackage{amsmath}	
\usepackage{amssymb}	
\usepackage{comment}
\usepackage{soul}
\usepackage{color}
\usepackage{xcolor}
\usepackage{hyperref}
\usepackage{multirow}

\def\msun{{\rm M_{\odot}}}
\newcommand\rsun{{\rm R_{\odot}}}

\def\be{\begin{equation}}
\def\ee{\end{equation}}

\def\del#1{{}}

\newcommand\MSunPerYear{~${\rm M_{\odot}}$~yr$^{-1}$\,}

\begin{document}


   \title{The Role of Thermal Instability in Accretion Outbursts in High-Mass Stars}


   \author{ Vardan G. Elbakyan\inst{1},
           Sergei Nayakshin\inst{2}, Alessio Caratti o Garatti\inst{3}, Rolf Kuiper\inst{1}, Zhen Guo\inst{4,5,6}
          }
   \institute{Fakult{\"a}t f{\"u}r Physik, Universit{\"a}t Duisburg-Essen, Lotharstra{\ss}e 1, D-47057 Duisburg, Germany; 
   \email{vardan.elbakyan@uni-due.de} 
    \and
     School of Physics and Astronomy, University of Leicester, Leicester, LE1 7RH, UK;  
    \and
    INAF-Osservatorio Astronomico di Capodimonte, Salita Moiariello 16, 80131 Napoli, Italy;
    \and
    Instituto de Física y Astronomía, Universidad de Valparaíso, ave. Gran Bretaña, 1111, Casilla 5030, Valparaíso, Chile;
    \and
    Centre for Astrophysics, University of Hertfordshire, College Lane, Hatfield AL10 9AB, UK;
    \and
    Millennium Institute of Astrophysics, Nuncio Monse{\~n}or Sotero Sanz 100, Of. 104, Providencia, Santiago, Chile
}

  \date{Received August XX, 2024; accepted November YY, 2024}
   
   \titlerunning{Accretion Outbursts in High-Mass Stars}
   \authorrunning{Elbakyan et al.}

  \abstract
  {High-mass young stellar objects (HMYSOs) can exhibit episodic bursts of accretion, accompanied by intense outflows and luminosity variations. Understanding the underlying mechanisms driving these phenomena is crucial for elucidating the early evolution of massive stars and their feedback on star formation processes.}
  {Thermal Instability (TI) due to Hydrogen ionisation is  among the most promising mechanisms of episodic accretion 
  in low mass ($M_*\lesssim 1\msun$) protostars. Its role in HMYSOs has not yet been elucidated. Here, we investigate the properties of TI outbursts in young, massive ($M_*\gtrsim 5$~$\msun$) stars, and compare them to those observed so far.}
  {We employ a 1D numerical model to simulate TI outbursts in HMYSO accretion disks. We vary key model parameters, such as stellar mass, mass accretion rate onto the disc, and disc viscosity, to assess TI outburst properties.}
  {
  Our simulations show that modelled TI bursts can replicate the durations and peak accretion rates of long (a few years to decades) outbursts observed in HMYSOs with similar mass characteristics. However, they struggle with short-duration (less than a year) bursts with short (a few weeks or months) rise times, suggesting the need for alternative mechanisms. Moreover, while our models match the durations of longer bursts, they fail to reproduce the multiple outbursts seen in some HMYSOs, regardless of model parameters. We also emphasise the significance of not just evaluating model accretion rates and durations, but also performing photometric analysis to thoroughly evaluate the consistency between model predictions and observational data. 
  }
  {Our findings suggest that 
  some other plausible mechanisms, such as gravitational instabilities and disc fragmentation can be responsible for generating the observed outburst phenomena in HMYSOs and underscore the need for further investigation into alternative mechanisms driving short outbursts. 
  However, the physics of TI is crucial in sculpting the inner disc physics in the early bright epoch of massive star formation, and comprehensive parameter space exploration and the use of 2D modeling are essential for obtaining a more detailed understanding of the underlying physical processes. By bridging theoretical predictions with observational constraints, this study contributes to advancing our knowledge of HMYSO accretion physics and the early evolution of massive stars. 
  }


   \maketitle
%

\section{Introduction}

Understanding the formation and evolution of high-mass young stellar objects (HMYSOs), and their various accretion outbursts, requires a comprehensive study that includes observations, analytical models, and numerical simulations. HMYSOs hold crucial information on the formation process of massive stars, which play a key role in shaping the cosmic environment. 
While FU Orionis-type accretion outbursts are typically associated with low-mass young stellar objects \citep{1996HartmannKenyon}, recent observations have suggested that HMYSOs may also experience similar episodic accretion events \citep{2017Caratti, 2021Stecklum, 2022Fischer}. However, the mechanisms driving these outbursts in HMYSOs are likely different, given the distinct physical conditions in high-mass star formation.

FU Orionis type accretion outbursts are a puzzling and fascinating phenomenon in astrophysics, primarily observed in low-mass young stellar objects (YSOs). These outbursts, lasting for several years or decades, involve an intense increase in the accretion rate onto a young star, leading to a brightening of several magnitudes. The prototype of this class of outbursts is FU Orionis itself, a low-mass star discovered more than 85 years ago \citep{1977Herbig}. Since then, many other examples of FUors have been identified, in different regions of our galaxy \citep{AudardEtal14}, providing evidence that these phenomena might be ubiquitous in low-mass star-forming environments. Later on, with the discovery of EXor bursts \citep{Herbig89}, it was realized that episodic accretion is a much broader and varied phenomenon, that occurs at all stages of star formation. The outburst of V1647 Ori in 2004 showed that some objects do not fit into the traditional EXor and FUor categories, despite the broad range of properties these groups cover. Properties of  several recently discovered outbursting objects are even more unusual, thus creating a zoo of outbursting objects with various properties \citep[see][for the review]{2022Fischer}.

Despite several attempts to explain the physical processes driving these outbursts, the exact mechanism behind them remains unclear. Nevertheless, numerous theoretical scenarios have been proposed, ranging from gravitational instability (GI) in the disc around the young star \citep{2010VorobyovBasu, 2011MachidaInutsuka, 2016Hosokawa, 2017MeyerVorobyov, 2018VorobyovElbakyan, 2018ZhaoCaselli, 2018KuffmeierFrimann, 2020OlivaKuiper}, thermal instability (TI) \citep{1994BellLin, 1996KleyLin}, stellar flyby \citep{2022Borchert, 2023Cuello}, interaction of a massive planet with the disc \citep{2004LodatoClarke} or the presence of a "dead zone" in the inner $0.1 \leq r \leq$~a~few~AU disc with eventual development of the magnetorotational instability (MRI) in that zone \citep[][]{2001Armitage, 2009ZhuHartmann, 2010ZhuHartmann, 2020Kadam, 2020VorobyovKhaibrakhmanov}.

One of the possible mechanisms responsible for the episodic accretion outbursts is the well-established phenomenon of disc thermal instability caused by the sudden increase in opacity when hydrogen ionizes at a temperature of $\sim$$10^4$~K. The inner, thermally unstable region of the disc periodically switches between two (cold and hot) stable branches of the so-called S-curve \citep[see Sect.~6 in][]{2015Armitage}, thus developing a strong accretion variability onto the star \citep[e.g.,][]{2004LodatoClarke,2021ElbakyanNayakshin}. This phenomenon was originally studied as part of a model for dwarf novae \citep{1979Hoshi, 1981Meyer, 1985Lin}. This model, while proving to be a good fit to observations, requires not only a substantial change in opacity and distinct values of the $\alpha$-parameter for the cold and hot branches of the S-curve, but also that the disc midplane temperatures exceed $\sim$$10^4$~K.

Initial studies of accretion outbursts were focused mainly on a FU Orionis objects (FUors), which are low-mass stars exhibiting outbursts every few decades \citep{1996HartmannKenyon}. However, with the advent of observational techniques and new instruments (such as ALMA, VLTI, VISTA), several groups have discovered multiple HMYSOs showing accretion outbursts \citep{2017Caratti, 2018Hunter, 2021Stecklum, 2021Chen, 2024WolfStecklum}, which introduce some discrepancies among their observational properties compared to the FUors. These discrepancies could potentially help to fill in critical details of the underlying physical mechanisms at work, as well as refining theoretical predictions. While in low-mass YSOs we distinguish the two large categories as magnetospheric-accretion driven (EXors) and boundary layer accretion driven (FUors), in HMYSOs the situation is more complex, as in principle, magnetospheric accretion should not work \citep{2016Beltran, 2020Mendigutia}. The outbursts so far observed in HMYSOs are rapid and short-lived events, that are manifested by a sudden increase in luminosity that may last for months to a few decades before returning to quiescence. Unfortunately, due to the low statistics for the outbursts in HMYSOs (less than 10 objects), it's hard to determine their recurrent rate like it is done for FUors ($\sim1.75$~kyr; \citet{2024ContrerasPena}). It is worth to note that bursts similar to the one in FU Orionis system have not yet been detected in HMYSOs (all bursts observed so far have relatively short durations and low amplitudes compared to FU Ori bursts). However, the lack of detections is likely due to an observational bias (and the relative low number of such events), and we should keep an open mind and ask ourselves if FUor events are possible in HMYSOs. While  the observed sample is still too small to judge whether these short bursts represent a common and unique feature or are just caused by an observational bias, we can still draw some conclusions.

The short durations of the bursts correspond to the dynamical and viscous timescales at the (sub-)AU radial distances, meaning that only the region of the disc close to the star is involved in the burst triggering. However, the multidimensional numerical simulation of (sub-)AU radial distances in a disc is usually too computationally costly and a so-called “sink cell” approach is used to simulate the inner region of the disc \citep{2017MeyerVorobyov, 2020OlivaKuiper} or these small-scale structures are hidden within a Gaussian accretion kernel around sink particles \citep{2004Krumholz, 2020Mignon-Risse, 2022Commercon}. This approach does not allow simulating properly the hydrodynamical evolution of the inner disc, which is crucial for the understanding of accretion outbursts. The most practical way to overcome this "computational barrier" is the usage of one dimensional (1D) numerical models of circumstellar discs.

In our current study, we aim to advance our understanding of accretion outbursts in HMYSOs by investigating the main characteristics of the phenomenon and its fundamental effects on forming massive stars and their surrounding discs. This is done by studying the evolution of circumstellar discs around HMYSOs, which are prone to TI in the inner disc. 

In Sec.~\ref{sec:methods}, we detail the numerical methodology employed in this study. We discuss the implementation of classic thermal instability (TI) bursts in Sec.~\ref{sect:TI}, where we explore the fundamental characteristics of these bursts under varying model parameters. In Sec.~\ref{sec:TI_typical_s_curve}, we extend our analysis to investigate TI bursts for different sets of model parameters, systematically examining their impact on the observed burst behaviour of HMYSOs. Following this, in Sec.~\ref{sec:obser}, we compare our simulation results with observational data. In Sec.~\ref{sec:multiplicity}, we delve into the phenomenon of thermal instability bursts with higher frequency. The discussion section (Sec.~\ref{sec:discussion}) covers comparisons with observational data, possible burst mechanisms, and the importance of parameter space exploration. We conclude with our main findings and suggestions for future research in Sec.~\ref{sec:conclusion}.

\section{Numerical methodology}\label{sec:methods}

The evolution of the protoplanetary disc is simulated using a 1D Shakura-Sunyaev viscous disc model \citep{1973ShakuraSunyaev}, implemented in the hydrodynamical code DEO, with azimuthal symmetry and vertical averaging \citep{2012NayakshinLodato, 2021ElbakyanNayakshin, 2022NayakshinElbakyan}. The detailed description of the DEO model can be found in \citet{2023NayakshinElbakyan}, while here we present only the modified model properties. The master equation for the viscous evolution can be presented as follows:
\begin{equation}
\begin{split}
    \frac{\partial\Sigma}{\partial t} = \frac{3}{r} \frac{\partial}{\partial r} \left[ r^{1/2} \frac{\partial}{\partial r} \left(r^{1/2}\nu \Sigma\right) \right] + \\  \frac{\dot{M}_{\mathrm{dep}}}{2\pi^{3/2} r \sigma} \exp\left[-\left(\frac{r-r_{\mathrm{dep}}}{\sigma}\right)^2\right],
\end{split}
\label{dSigma_dt}
\end{equation}
where $\Sigma$ is the gas surface density, $\nu$ is the kinematic viscosity of the disc, and the last term on r.h.s. is responsible for the mass deposition from outer regions into the disc. The mass is deposited from the envelope into the disc at a constant rate, $\dot M_{\rm dep}$, in a Gaussian ring with a centroid radius $r_{\mathrm{dep}}=18$~AU and dispersion $\sigma=0.05 r_{\mathrm{dep}}$. Our computational grid is logarithmic, with default 124 radial zones covering the radial region from $R_{\rm in} = 0.1$~AU to 20~AU. The disc in our model is initially in a steady-state configuration, with an accretion rate $\dot M = \dot M_{\rm dep}$ at all disc radii.

In our models, we simulate only the inner 20 AU of the disc, while observed disc sizes around HMYSOs typically range from a few hundred to a few thousand AU \citep{2016Beltran}. Our simulated disks have masses ranging from a few times $10^{-1}$ $\msun$ to a few $\msun$, depending on the stellar mass, consistent with observations of circumstellar disks in HMYSOs \citep[e.g.,][]{2006Cesaroni}. This results in disc-to-star mass ratios greater than 0.1, which aligns with observational findings \citep[e.g.,][]{2023Ahmadi}. 

We parametrise the kinematic viscosity in the disc through the dimensionless $\alpha$ parameter, $\nu=\alpha c_{\rm{s}} H$ \citep{1973ShakuraSunyaev}, where $c_{\rm{s}}$ is the gas sound speed in the disc midplane and $H$ is the vertical scale height of the disc. 
Following \citet{1998Hameury}, we define the $\alpha$ parameter as 
\begin{equation}
    \ln\alpha = \ln\alpha_{\rm cold}\, + \,\frac{\ln\alpha_{\rm hot}-\ln\alpha_{\rm cold}}{1 + (T/T_{\rm cr})^8}
    \label{alpha_vs_T}
\end{equation}
where $T_{\rm cr}$ is a critical temperature (see below), $\alpha_{\rm cold}$ and $\alpha_{\rm hot}$ are $\alpha$-parameters on the cold and hot branches of the S-curve. Recent MHD simulations suggest $\alpha_{\rm cold}\approx10^{-2}$, while there is a large discrepancy for the values of $\alpha_{\rm hot}$.  Most of the global disc simulations adopt a magnetic field geometry that is similar to zero net-flux shearing-box and produce values of $\alpha_{\rm hot}\sim0.01-0.02$ \citep{2000MillerStone, 2006Hirose, 2006FromangNelson, 2010Shi, 2010Davis, 2011Flock, 2011Beckwith, 2012Simon}. Meanwhile, the value of $\alpha_{\rm hot}$ estimated from the observations of dwarf novae and X-ray binaries, found to be of the order of 0.1 \citep{1999Smak, 2001Dubus}. It is therefore questioned, how correct it is to use zero mean field models for simulations of accretion discs \citep[see Discussions in][]{KingEtal07, 2009ZhuHartmann, BaiStone13}. Recent 3D MHD models show that the joint action of convection and magnetorotational instability can increase $\alpha_{\rm hot}$ to the observational values \citep{2014Hirose, 2018Coleman, 2018Scepi}. 
The values of $\alpha_{\rm cold} = 0.01$ and $\alpha_{\rm hot} = 0.1$ were found to work relatively well for both observations \citep[e.g.,][]{2001Lasota} and first-principle simulations of TI in discs \citep[e.g.,][]{2014Hirose}.

Thermal equilibrium in protoplanetary disks is crucial for understanding the temperature distribution and physical conditions within these disks. It relies on the equilibrium between the heating and cooling processes within the disc's vertical structure. In our previous works \citep{2023NayakshinOwen, 2023NayakshinElbakyan, 2023ElbakyanNayakshin}, we used a time-dependent energy balance equation with a one-zone approximation for the vertical thermal equilibrium and a radial advection of the heat \citep{2004LodatoClarke, 2012NayakshinLodato}.
Such approach is a useful tool for providing a simplified overview of the thermal balance in protoplanetary disks. However, it should be used with caution and an awareness of its limitations. 

For a more accurate representation of the complex physical processes occurring in disks, we improved the energy balance equation in our model, following \citet{1993Cannizzo, 1999Menou}. Instead of the one-zone approximation in the vertical direction, we solve the z-dependent vertical balance equations (cf. Sect.~2.3 in \citet{2024Nayakshin}) 

\begin{equation}
    \frac{dP(z)}{dz} = -\Omega^2\rho(z)z,
\end{equation}
where $P(z)$ and $\rho(z)$ are the local gas pressure and density, respectively. We find P for a given $\rho$ and $T$, where the temperature $T$ satisfies the equation
\begin{equation}
    \frac{dT(z)}{dz} = -\frac{3\kappa(\rho,T)\rho F(z)}{16\sigma_{\rm B}T^3},
\end{equation}
where 
\begin{equation}
    F(z) = F_{\rm rad}\frac{z}{z_{2/3}} = \sigma_{\rm B} T_{\rm eff}^4\frac{z}{z_{2/3}}
\end{equation}
where $z_{2/3}$ is the location of the $\tau(z) = 2/3$ surface with the temperature $T(z) = T_{\rm eff}$. 

The disc temperature is evolving as
\begin{equation}
    \frac{\partial T}{\partial t} = - \frac{T-T_{\mathrm{eq}}}{t_{\mathrm{th}}} - \frac{1}{r} \frac{\partial}{\partial r} (Trv_{r}) +  \frac{3\nu}{r} \frac{\partial}{\partial r} \left( r\frac{\partial T}{\partial r} \right)
\end{equation}
where $t_{\mathrm{th}}$ is the thermal timescale, $v_r$ is the radial velocity and $T_{\mathrm{eq}}$ is the equilibrium temperature in the disc at the radial distance $r$, found from the assumption of the vertical energy balance for the disc. In the one-zone approximation, it is assumed that the opacity $\kappa$ in the disc is the same at the midplane and at its surface. However, $\kappa$ strongly depends on the temperature in the disc, which can vary by up to an order of magnitude between the midplane and the disc surface. The z-dependent vertical balance equations eliminate the need to assume that $\kappa$ in the disc is the same in the vertical direction. In our model, we use opacities from \citet{2009ZhuHartmann}. A more detailed description of the model can be found in \citet{2024Nayakshin}. For the models presented in this section, we set $T_{\rm cr}=8000$~K, unlike the $T_{\rm cr}=2.5\times 10^4$~K used for the one-zone models earlier. The reasons for it and the models with higher $T_{\rm cr}$ are discussed in Sect.~\ref{sec:multiplicity}.

\section{Classic Thermal Instability bursts}
\label{sect:TI}

In this section, we study classic thermal instability (TI) bursts in HMYSO discs. We aim to establish their main characteristics, such as mass accretion rate during the burst, its rise time, duration, etc. We also explore how parameters of the problem, such as $M_*$, viscosity parameter in the hot state ($\alpha_{\rm hot}$) and feeding rate, affect the observability of the bursts.

\begin{figure}
\begin{centering}
\includegraphics[width=1\columnwidth]{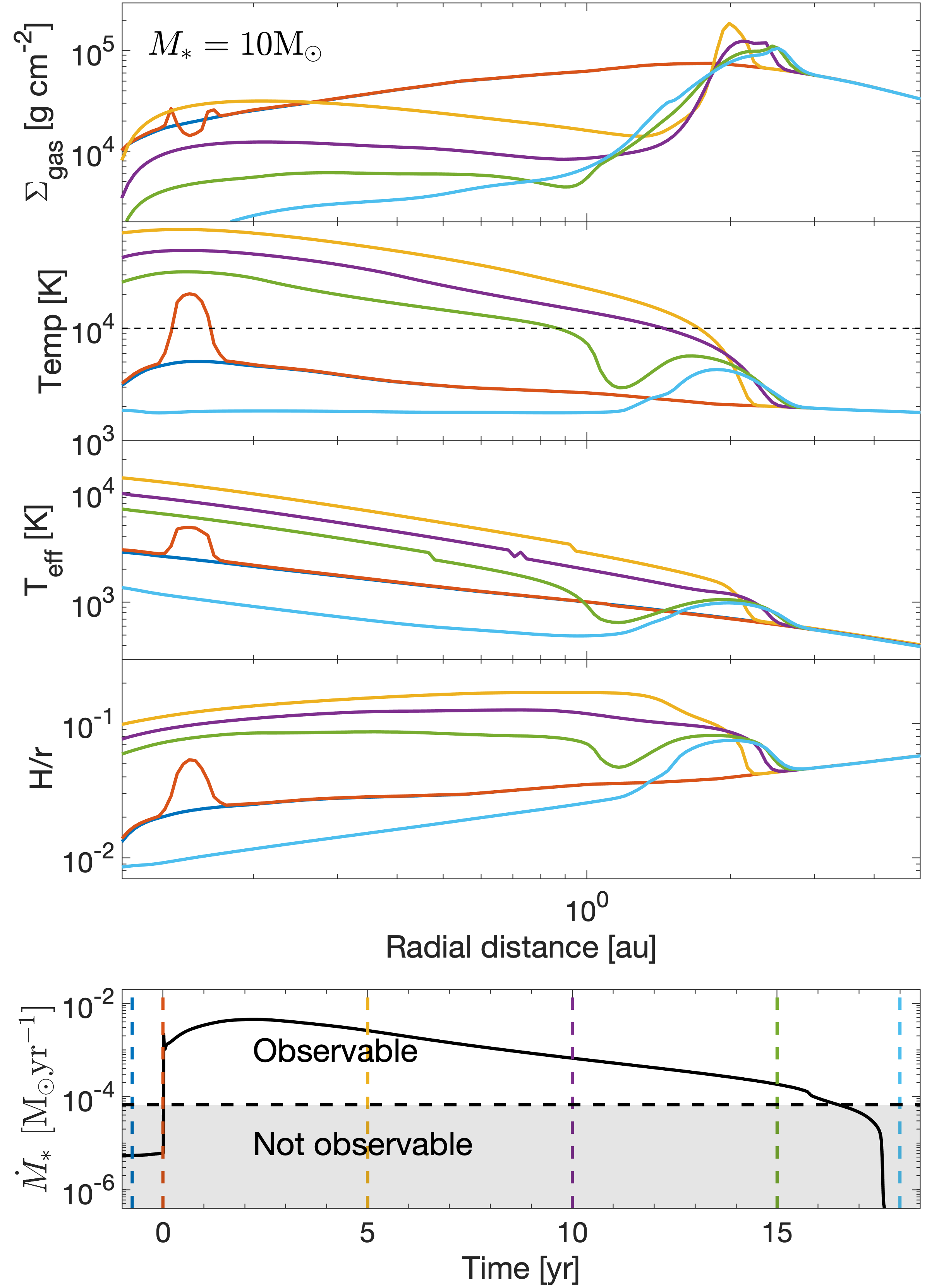}
\par\end{centering}
\caption{\label{fig:radial_profiles} \textbf{Top four panels:} The radial profiles of gas surface density (top panel), midplane disc temperature (second panel), effective disc temperature (third panel), and disc aspect ratio (fourth panel) for five distinct time instances. \textbf{Bottom panel:} Mass accretion rate history onto the HMYSO during the TI outburst. Vertical dashed lines indicate the time instances for which the radial profiles with corresponding colours are shown on the top panels. The horizontal dashed line shows the value of $\dot M_{\rm th}$ calculated with Eq.~(\ref{eq:mdot_obs}). The shaded area shows the non-observable part of the accretion rate variations.}
\end{figure}

We start by considering and briefly describing one thermal instability cycle for a model with $M_* = 10 \, \msun$, $\dot M_{\rm dep} = 10^{-4}~\msun$~yr$^{-1}$, and the value of $\alpha_{\rm cold} = 0.01$ and  $\alpha_{\rm hot} = 0.1$. 
The solid black curve in the bottom panel of Figure~\ref{fig:radial_profiles} shows the mass accretion rate onto the star vs time for this model. The top four panels of Figure~\ref{fig:radial_profiles} present the radial profiles of key disc variables -- surface density $\Sigma$, midplane temperature, effective temperature, and disc aspect ratio $H/r$ -- at five distinct time instances indicated with the corresponding vertical lines in the bottom panel.  
The time $t=0$ corresponds to the very beginning of the burst. As it is well known, classic TI bursts are triggered very close to the disc inner boundary and propagate outward \citep[e.g.,][]{1994BellLin}. This is born out by our models: the temperature jumps to the hot stable branch at $r\approx 0.14$~AU first, where hydrogen becomes ionized. The ionisation front then propagates both inward and outward. The inward propagating front produces the $\dot M\approx 10^{-3}$\MSunPerYear burst onto the star, with the rise time of $\approx0.1$~yr, which is sustained for less than a year (a fraction of the local viscous time). Subsequently, the ionisation front propagating outward puts the disc onto the hot stable branch at increasingly larger radii but $\dot M_*$ keeps monotonically decreasing with time. Eventually the front stalls at $r_{\rm TI}\approx 2$~au and then retreats back as the disc falls into quiescence once again. The value of $r_{\rm TI}$ can be found semi-analytically as \citep{1994BellLin, 2024Nayakshin}

\begin{equation}
    r_{\mathrm{TI}} = 20\rsun\left(\frac{\dot{M}_{\mathrm{dep}}}{3\times10^{-6}}\right)^{1/3}\left( \frac{M_*}{\msun} \right)^{1/3}  \left( \frac{T_{\mathrm{eff}}}{2000} \right)^{-4/3}
    \label{eq:r_TI}
\end{equation}

Thus, the region of the disc influenced by thermal instability is limited to where the effective temperature in the steady-state disc exceeds approximately 2000~K. As shown in the fourth panel of Figure~\ref{fig:radial_profiles}, our model aligns well with this analytical estimate, with the ionization front halting roughly at the point where the disc's effective temperature falls below 2000~K.

Due to the inside-out propagation of the ionisation front, a region with increased density/temperature is forming at the location where the front stalls. If the density/temperature in this region becomes critical to trigger TI, a new, so-called reflare, will be initiated at this location of the disc \citep{2000Menou, 2001Dubus}. We discuss models with reflares in Sect.~\ref{sec:reflares}.
At the end of the outburst, the
surface density profile of the disc does not recover its pre-burst configuration (compare the dark blue and light blue curves in the top panel). This is due to the disc being emptied out onto the star during the burst. The next outburst will repeat only when the inner disc region is resupplied with fresh material from outside. The duration of the burst is of the order of the disc viscous time at the outer radius of the ionised zone, $t_{\rm visc}=r_{\rm TI}^2/\nu$. 

One important observational difference between TI outbursts  on low-mass and high-mass stars is the fact that the latter are orders of magnitude brighter than the former. TI outbursts must outshine the star to be observationally important. A crude but useful precondition for a burst to be observable is for the bolometric accretion luminosity, $L_{\rm acc}=  G M_* \dot M_*/R_*$, to exceed the pre-burst luminosity of the central star, $L_*$. We hence introduce the threshold value $\dot M_{\rm th}$, and assume that a TI burst is observable when the mass accretion rate onto the star is higher than this threshold value, defined as
\begin{equation}
    \dot{M}_{\rm th}=\frac{L_*R_*}{GM_*},
    \label{eq:mdot_obs}
\end{equation}
where $L_*=1.4(M_*/\msun)^{3.5} L_{\odot}$ is the stellar luminosity, $R_*=(M_*/\msun)^{0.7} R_{\odot}$ is the stellar radius and $M_*$ is the stellar mass. 
Here we used a mass-luminosity and mass-radius relations for  ZAMS stars and this threshold value can be considered as a conservative minimal value, since the changes of stellar luminosity and radius during the intensive mass accretion are neglected. During the accretion outburst, with the high accretion rates of $\dot M_*>10^{-4}~\msun$~yr$^{-1}$, the radius of HMYSO can change dramatically, reaching a few hundred solar radii \citep{2009HosokawaOmukai, 2013KuiperYorke, 2017TanakaTan, 2019MeyerVorobyov}, while during more moderate accretion with $\dot M_*<10^{-4}~\msun$~yr$^{-1}$ the stars with masses $M\gtrsim10$~$\msun$ have radii close to the ZAMS one. For our models, to avoid discrepancy and addition of new free parameters, we set $R_*\approx20$~$\rsun$ for all our models, which is an intermediate value between the two extreme cases. However, for the calculation of $\dot{M}_{\rm th}$, we use the $R_* 
\propto M_*^{0.7}$ relation, 
since it is a more conservative value and cannot lead to the underestimation of the number or duration of TI burst.  We plan to conduct a study with the self-consistent evolution of HMYSO radius using a stellar evolution code in the future. The horizontal line in the bottom panel of Figure~\ref{fig:radial_profiles} shows $\dot{M}_{\rm th}$ in our model.

Further, we discuss in detail the burst duration in our model with $\alpha_{\rm hot}=0.1$ and discuss the influence of $\alpha_{\rm hot}$ parameter on the burst observability and duration.

\section{TI bursts for different model parameters}\label{sec:TI_typical_s_curve}

\begin{figure}
\begin{centering}
\includegraphics[width=1\columnwidth]{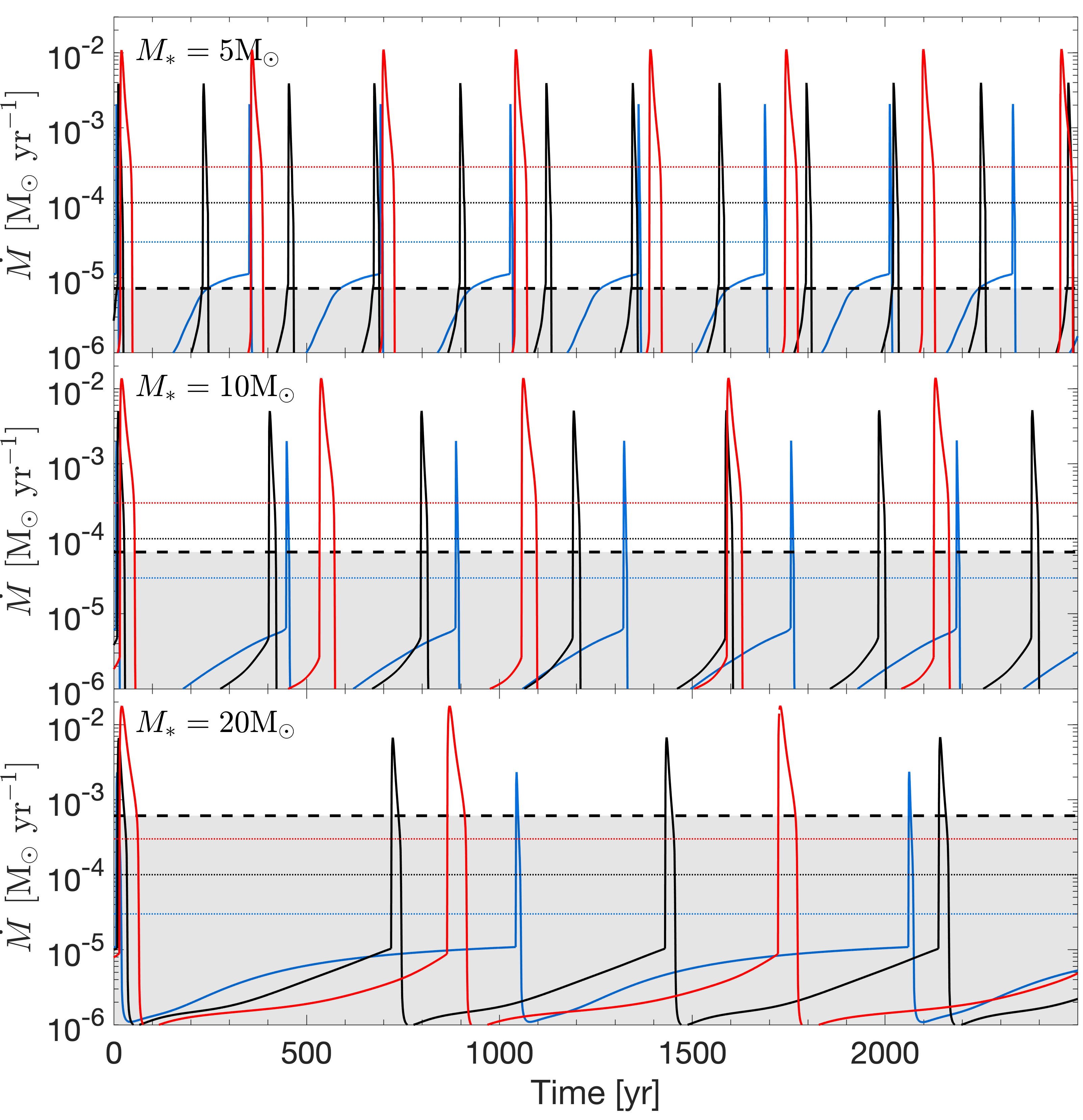}
\par\end{centering}
\caption{\label{fig:mdot_all_masses_long} Accretion rate onto the star of mass $M_* = 5$~$\msun$ (top panel), $M_* = 10$~$\msun$ (middle panel), and $M_* = 20$~$\msun$ (bottom panel) during TI bursts for three different values of $\dot M_{\rm dep}$ -- $3\times10^{-5}$~$\msun$~yr$^{-1}$ (blue line), $10^{-4}$~$\msun$~yr$^{-1}$ (black line), and $3\times10^{-4}$~$\msun$~yr$^{-1}$ (red line). The horizontal black dashed line shows the $\dot M_{\rm th}$ value for each stellar mass calculated with Eq.~(\ref{eq:mdot_obs}). The accretion rate variability below the dashed line (the shaded region) is unlikely to be observable. The horizontal dotted lines show the $\dot M_{\rm dep}$ for each model with the corresponding colour. The time instance $t=0$ in all panels of the figure is arbitrary. }
\end{figure}

\begin{figure}
\begin{centering}
\includegraphics[width=1\columnwidth]{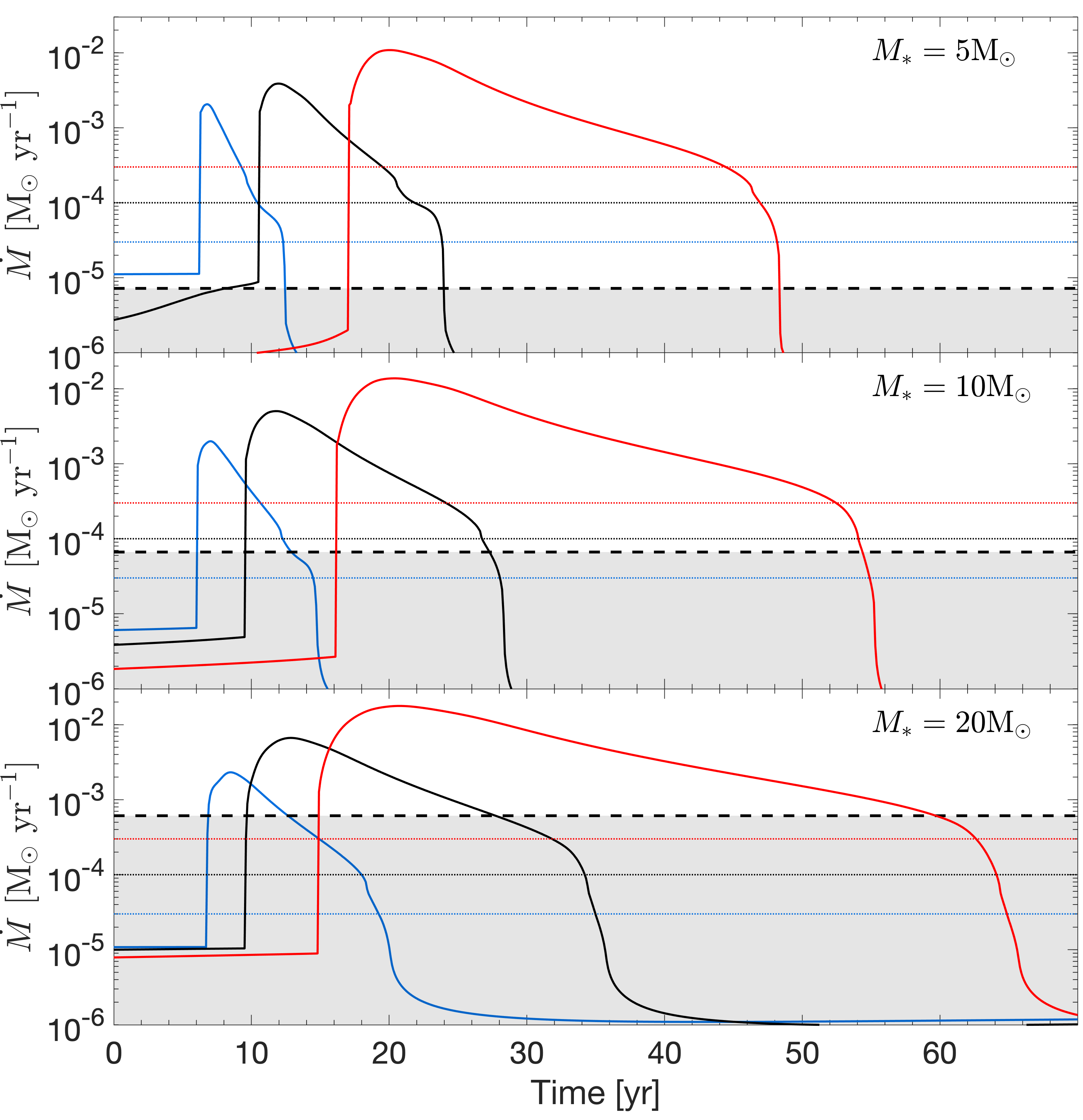}
\par\end{centering}
\caption{\label{fig:mdot_all_masses} Similar to Figure~\ref{fig:mdot_all_masses_long}, but zoomed-in on a single TI outburst in each model. }
\end{figure}

In Figure~\ref{fig:mdot_all_masses_long} we show long term (2500 years) temporal evolution of mass accretion rate onto the HMYSO for three values of stellar mass, $M_* = 5$~$\msun$ (top panel), $M_* = 10$~$\msun$ (middle panel), and $M_* = 20$~$\msun$ (bottom panel). The observed gas infall rates in high-mass star forming regions are estimated to range from $10^{-5}-10^{-3}$~$\msun$~yr$^{-1}$ \citep{2013Beuther, 2017Beuther, 2021Moscadelli}. Thus, for each $M_*$ we calculate three models with different values of $\dot M_{\rm dep}$: $3\times10^{-5}$~$\msun$~yr$^{-1}$ (blue line), $10^{-4}$~$\msun$~yr$^{-1}$ (black line), and $3\times10^{-4}$~$\msun$~yr$^{-1}$ (red line). The threshold $\dot M_{\rm th}$ (Eq.~(\ref{eq:mdot_obs})) is shown with the horizontal dashed line. The variability of the accretion rate in  the shaded region below that line will be not visible. Clearly, the TI bursts for all stellar masses are observable. 

Figure~\ref{fig:mdot_all_masses_long} demonstrates that TI burst  duration, amplitude, and repetition timescale, $t_{\rm rep}$, all  depend on $\dot M_{\rm dep}$. We find that $t_{\rm rep}$ increases with $M_*$, being on average $\sim300$~years for $M_* = 5$~$\msun$, $\sim450$~years for $M_* = 10$~$\msun$, and $\sim850$~years for $M_* = 20$~$\msun$ models. This implies that classical TI bursts could happen multiple times on the HMYSOs during their formation timescale of a few $\times10^5$~yrs \citep{2021Sabatini}. It is interesting to note, that even with less than 10 HMYSO bursts observed so far (cf. Sect.~\ref{sec:obser}), there are objects with multiple bursts, for which $t_{\rm rep} \sim 6-30$~yrs. There may however be physics leading to bursts with far shorter $t_{\rm rep}$ (Sect.~\ref{sec:multiplicity}).

In a quasi-steady state, TI bursts are highly periodic and repeatable. Thus, we choose a single TI outburst from each simulation shown in Figure~\ref{fig:mdot_all_masses_long} to study its properties in more detail in Figure~\ref{fig:mdot_all_masses}. 

We first focus on models with $M_* = 5$~$\msun$ (top panel).
The peak accretion rates during TI bursts for $\dot M_{\rm dep}=3\times10^{-5}$~$\msun$~yr$^{-1}$ and $\dot M_{\rm dep}=10^{-4}$~$\msun$~yr$^{-1}$ are $2\times10^{-3}$~$\msun$~yr$^{-1}$ and $3.9\times10^{-3}$~$\msun$~yr$^{-1}$, correspondingly. These peak values are comparable to those inferred in observations, $\dot M_*\approx10^{-3}~\msun$~yr$^{-1}$ (see Table~\ref{tab:1}). The model with $\dot M_{\rm dep}=3\times10^{-4}$~$\msun$~yr$^{-1}$ has a peak value of $10^{-2}$~$\msun$~yr$^{-1}$, which is 1 order of mag higher than the observed ones. All three models show a sharp rise at the beginning of the outburst, which then slows somewhat before the maximum is reached, followed by a yet slower decay. The burst rise timescales are somewhat consistent with the observations, being mostly longer. The duration of the observable part of the burst for $\dot M_{\rm dep}=3\times10^{-5}$~$\msun$~yr$^{-1}$, $\dot M_{\rm dep}=10^{-4}$~$\msun$~yr$^{-1}$, and $\dot M_{\rm dep}=3\times10^{-4}$~$\msun$~yr$^{-1}$ models are $\approx 6.2$, 13.5, and 31.3 years, respectively, longer than many of the bursts observed so far.

The models for $M_* = 10\, \msun$ and $M_* = 20\, \msun$ are shown in the middle and bottom panels of Figure~\ref{fig:mdot_all_masses}, respectively. 
The peak accretion rates during the TI bursts in all three models with $ M_* = 10\, \msun $ are slightly higher compared to the $ M_* = 5\, \msun $ case, which contributes to longer burst durations. The burst durations are $\approx 7$, 17.8, and 38.2 years for the models with $ \dot{M}_{\rm dep} = 3 \times 10^{-5} \, \msun \, {\rm yr}^{-1} $, $ \dot{M}_{\rm dep} = 10^{-4} \, \msun \, {\rm yr}^{-1} $, and $ \dot{M}_{\rm dep} = 3 \times 10^{-4} \, \msun \, {\rm yr}^{-1} $, respectively. In these cases, the longer burst durations are due to the thermal instability being triggered at larger radii in the disc, resulting in a greater portion of the disc contributing to the accretion process.
We note that the burst durations are longer, even considering the fact that $\dot{M}_{\rm th}$ for $M_* = 10\, \msun$ star is higher than for $M_* = 5\, \msun$ star. 

For the models with $M_* = 20\, \msun$ star, the peak accretion rates are higher than for $M_* = 10\, \msun$ case being $2.5\times10^{-3}$~$\msun$~yr$^{-1}$, $6.6\times10^{-3}$~$\msun$~yr$^{-1}$, and $1.8\times10^{-2}$~$\msun$~yr$^{-1}$, correspondingly, for $\dot M_{\rm dep}=3\times10^{-5}$~$\msun$~yr$^{-1}$, $\dot M_{\rm dep}=10^{-4}$~$\msun$~yr$^{-1}$, and $\dot M_{\rm dep}=3\times10^{-4}$~$\msun$~yr$^{-1}$ models. However, due to the higher $\dot{M}_{\rm th}$, not all the models have longer durations compared to the $M_* = 10\, \msun$ models. The burst durations are $\approx 6$, 18, and 45 years, respectively. Thus, only the burst in the model with the highest $\dot M_{\rm dep}$ becomes longer, while the burst in the model with the lowest $\dot M_{\rm dep}$ becomes shorter. We note that due to the higher $\dot{M}_{\rm th}$, most of the initial sharp rise with the rise-time of $<1$~yr is not observable and the amplitudes of the observable parts of the bursts become lower.

The model burst durations for all the considered stellar masses are consistent with some of the bursts observed in the HMYSOs with similar stellar mass (see Sect.~\ref{sec:obser}). However, the models are unable to reproduce the observed multiple bursts in HMYSOs or bursts with short durations ($<$~a~few~years). Multiple outbursts are discussed in more detail further in Sect.~\ref{sec:multiplicity}. The properties of the above-mentioned model bursts are summarized in Table~\ref{tab:sum_table}.

\begin{table}[]
    \centering
    \begin{tabular}{ccccc}
    \hline 
    \hline 
    Mass & $\dot{M}_{\rm dep}$ & $\dot{M}_{\rm peak}$ & Burst duration & $t_{\rm rep}$ \tabularnewline
    $[M_{\odot}]$ & $[M_{\odot} \rm{yr}^{-1}]$ & $[M_{\odot} \rm{yr}^{-1}]$ & [yr] & [yr] \tabularnewline
    \hline 
    \multirow{3}{*}{5} & $3\times10^{-5}$ & $2\times10^{-3}$ & 6.2 & 325 \tabularnewline
    & $10^{-4}$ & $3.9\times10^{-3}$ & 13.5 & 225 \tabularnewline
    & $3\times10^{-4}$ & $10^{-2}$ & 31.3 & 345 \tabularnewline
\tabularnewline
    \multirow{3}{*}{10} & $3\times10^{-5}$ & $2\times10^{-3}$ & 7 & 440 \tabularnewline
    & $10^{-4}$ & $5\times10^{-3}$ & 17.8 & 395 \tabularnewline
    & $3\times10^{-4}$ & $1.4\times10^{-2}$ & 38.2 & 520 \tabularnewline
\tabularnewline
    \multirow{3}{*}{20} & $3\times10^{-5}$ & $2.5\times10^{-3}$ & 6 & 1020 \tabularnewline
    & $10^{-4}$ & $6.6\times10^{-3}$ & 18 & 710 \tabularnewline
    & $3\times10^{-4}$ & $1.8\times10^{-2}$ & 45 & 850 \tabularnewline
    \hline
    \end{tabular}
    \caption{Burst properties in the models shown in Figure \ref{fig:mdot_all_masses_long} and \ref{fig:mdot_all_masses}.}
    \label{tab:sum_table}
\end{table}

\begin{figure}
\begin{centering}
\includegraphics[width=1\columnwidth]{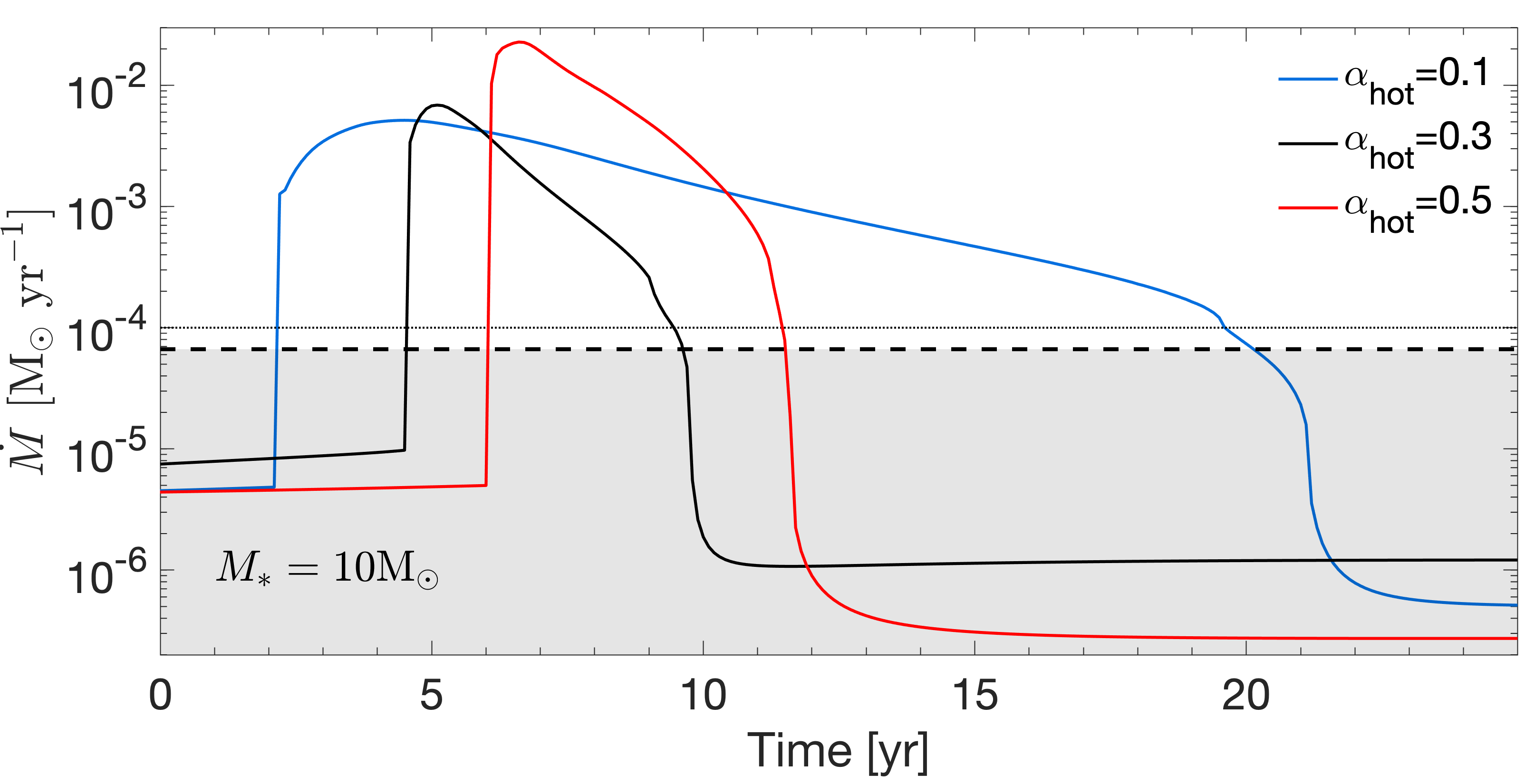}
\par\end{centering}
\caption{\label{fig:mdot_all_alphas} Accretion rate on the star of mass $M_* = 10$~$\msun$ during a single TI burst in the models with three different values of $\alpha_{\rm hot}$ -- 0.1 (blue line), 0.3 (black line), and 0.5 (red line). The horizontal dashed line shows the $\dot M_{\rm th}$ value calculated with Eq.~(\ref{eq:mdot_obs}). The accretion rate variability below the dashed line (the shaded region) is not observable.
 The horizontal dotted line shows the $\dot M_{\rm dep}$, which is equal to $\approx6.6\times10^{-5}$~$\msun$~yr$^{-1}$ for all the models. The time instance $t=0$ is arbitrary. }
\end{figure}

While the form of the turbulent viscosity parameter $\alpha$ introduced by \cite{1973ShakuraSunyaev} is constrained relatively well for TI bursts on solar mass objects \citep[see Sect.~\ref{sec:methods} and, e.g.,][]{2015Hirose,2018Scepi,Hameury-20-review}, the environment of HMYSO may differ. For that reason, here we explore how TI burst properties depend on the value of $\alpha_{\rm hot}$.
It is well known that the larger is the $\alpha$ value, the more efficient is the angular momentum transport in the disc, resulting in faster mass accretion towards the central object. In order to test, how the burst duration depends on the value of $\alpha_{\rm hot}$, we calculated two additional models of a thermally unstable disc around a star with mass 10~$\msun$. We assume $\dot M_{\rm dep}=10^{-4}$~$\msun$~yr$^{-1}$ for both models and employ $\alpha_{\rm hot}=0.3$ and $\alpha_{\rm hot}=0.5$. We note that such high values of $\alpha_{\rm hot}$ are not consistent with the value $\alpha_{\rm hot}\approx0.1$ estimated from the observations \citep[e.g.,][]{2001Dubus} and simulations \citep[e.g.,][]{2018Coleman, 2018Scepi}, but are considered here to push the model to the theoretical limits. In Figure~\ref{fig:mdot_all_alphas}, we show the accretion rate onto the HMYSO during a single TI outburst in each model, with a distinct value of $\alpha_{\rm hot}$ shown in the top right corner of the figure. The model with $\alpha_{\rm hot}=0.1$ (blue line) is similar to the one shown in the middle panel of Figure~\ref{fig:mdot_all_masses} (please note the difference in timescales). As expected, the burst durations are shorter in the models with higher $\alpha_{\rm hot}$. Interestingly, the burst durations show not much difference between the models with $\alpha_{\rm hot}=0.3$ (black line) and $\alpha_{\rm hot}=0.5$ (red line), being 5 and 5.5 years, respectively. On the other hand, the increase in $\alpha_{\rm hot}$ changes substantially the amplitude of the outburst. The accretion rate at the peak of the TI burst in $\alpha_{\rm hot}=0.3$ model is $7\times10^{-3}$~$\msun$~yr$^{-1}$, while in the $\alpha_{\rm hot}=0.5$ model is $2.3\times10^{-2}$~$\msun$~yr$^{-1}$. Thus, the increase in $\alpha_{\rm hot}$ from 0.1 to 0.3 shortens the duration of the burst. However, the further increase of $\alpha_{\rm hot}$ from 0.3 to 0.5 increases the amplitude of the burst and leads to a slight increase of the burst duration. As a result, even unrealistically high values of $\alpha_{\rm hot}$ are not able to shorten the duration of TI bursts in HMYSOs to less than 5 years and TI bursts can be responsible only for the bursts with longer durations. The burst properties in the models with various $\alpha_{\rm hot}$ are summarized in Table~\ref{tab:sum_table_alpha_hot}.

\begin{table}[]
    \centering
    \begin{tabular}{ccccc}
    \hline 
    \hline 
    Mass & $\alpha_{\rm hot}$ & $\dot{M}_{\rm peak}$ & Burst duration \tabularnewline
    $[M_{\odot}]$ &  & $[\times10^{-3} M_{\odot} \rm{yr}^{-1}]$ & [yr] \tabularnewline
    \hline 

    \multirow{3}{*}{10} & 0.1 & $5$ & 17.8 \tabularnewline
    & 0.3 & $7$ & 5 \tabularnewline
    & 0.5 & $23$ & 5.5  \tabularnewline

    \hline 
    \end{tabular}
    \caption{Burst properties in the models shown in Figure \ref{fig:mdot_all_alphas}.}
    \label{tab:sum_table_alpha_hot}
\end{table}

\section{Comparison with the observations} \label{sec:obser}

So far, only a few observed HMYSOs accretion bursts are characterised sufficiently for a meaningful comparison to our simulations; we list them in Table~\ref{tab:1}. Their durations range from a few months to at least a decade \citep{2017Caratti, 2017Hunter, 2019Brogan, 2019Proven-Adzri, 2021Stecklum, 2021Chen, 2024WolfStecklum}. Here we focus mainly on two properties of the bursts - the peak mass accretion rate during the burst, $\dot{M}_{\rm peak}$, and the duration of the burst. All the observed bursts on HMYSOs have $\dot{M}_{\rm peak}\gtrsim10^{-3}$~$\msun$~yr$^{-1}$. The observed short burst durations indicate that the processes leading to the bursts occur in the immediate vicinity to the central star \citep{2021ElbakyanNayakshin}.

\begin{table*}
\center
\caption{\label{tab:1} Main properties of outbursting HMYSOs.}
\begin{tabular}{cccccccc}
\hline 
\hline 

Object & Mass & Radius & Burst duration & Rise time & $\dot{M}_{\rm peak}$ & Accreted mass & Multiple \tabularnewline
& [$M_{\odot}$] & [$R_{\odot}$] & [yr] & [yr] & [$M_{\odot} \rm{yr}^{-1}$] & [$M_J$] & bursts  \tabularnewline
\hline 
M17 MIR$^{\S}$ & 5.4 & $10.1-34.1$ & $9-20$ & $<$1 & 5$\times10^{-3}$, 2$\times10^{-3}$ & $\sim$30($\times2$) & yes \tabularnewline
NGC 6334I MM1$^*$ & 6.7 & $2.6-20$ &  $>$8 & 0.62$^{+0.14}_{-0.14}$ & $2.3\times10^{-3}$ & 0.325$^{+0.126}_{-0.094}$& no\tabularnewline
G358.93-0.03 MM1$^\ddag$ & 9.7$^{+0.3}_{-0.6}$ & 3.9$^{+0.1}_{-0.2}$ & 0.75 & 0.16$^{+0.01}_{-0.01}$  & $1.8^{+1.2}_{-1.1}\times10^{-3}$ & 0.566 & no \tabularnewline
V723 Car$^{\P}$ & 10 & $18-58$ &  5 & 4 & - & - & no?\tabularnewline
S255IR NIRS 3$^\dagger$ & 20 & 10 &  $2-2.5$ & 0.4 & 5$\times10^{-3}$ & 2 & yes$^\$$ \tabularnewline
G323.46-0.08$^{**}$ & 23 & - & 8.4 & 1.4 & 0.8$\times10^{-3}$ \tabularnewline
W51 IRS2$^{\diamond}$ & 40-60$^{\triangledown}$ & - &  $<$1 & $\ll$1 & - & $\sim$260($0.25\msun$) & no\tabularnewline

\hline 
\end{tabular}
\\
\textbf{References.}
$\dagger$\citet{2017Caratti}, \;
$^*$\citet{2017Hunter, 2021Hunter}, \;
$\ddag$\citet{2021Stecklum}, \; \\
$^{\P}$\citet{2015Tapia}, \;
$^{**}$\citet{2019Proven-Adzri,2024WolfStecklum}, \;
$^{\S}$\citet{2021Chen}, \;
$^{\diamond}$\citet{2022Zhang}, \;
$^{\triangledown}$\citet{2009Zapata}. $^\$$Based on sparse photometric observations of \citet{2023Fedriani}.
\end{table*}

The star in M17 NIR is an intermediate mass one, $M_*\approx 5\, \msun$. The top panel in Figure~\ref{fig:mdot_all_masses} shows that classic TI bursts on such stars are observable for all parameter values considered, due to the relatively low value of threshold $\dot{M}_{\rm th}$. The TI bursts in the models with $\dot M_{\rm dep} = 3\times10^{-5}$~$\msun$~yr$^{-1}$ and $\dot M_{\rm dep} = 3\times10^{-4}$~$\msun$~yr$^{-1}$ are not consistent with the ones observed in M17 NIR, having either shorter (longer) durations or higher (lower) $\dot{M}_{\rm peak}$.
On the other hand, TI burst in the model with $\dot M_{\rm dep} = 10^{-4}$~$\msun$~yr$^{-1}$ has burst duration and $\dot{M}_{\rm peak}$ that are consistent with the observed burst in M17 NIR.
Thus,  TI may be responsible for an outburst on M17 MIR (HMYSO with mass $M\approx5$~$\msun$), if the burst had a duration of about a decade and is a singular outburst. However, as it was stated earlier, M17 MIR showed multiple outbursts, and we address this problem in Sect.~\ref{sec:multiplicity}.

Similar to the models with $M_*=5\, \msun$, all model TI outbursts on HMYSOs with mass $M_*=10$~$\msun$ are observable (see the middle panel in Fig.~\ref{fig:mdot_all_masses}). 
However, these  outbursts have durations of $\gtrsim7$~years, which is longer than the observed outbursts on the HMYSOs with the similar mass (0.75--5 years). Although the estimated mass of NGC 6334I MM1 is about 6.7~$\msun$ \citep{2021Hunter}, the burst duration in this HMYSO has a duration of more than 8 years, resembling the model outburst duration for $M_*=5\, \msun$ star. 

We arrive at a similar conclusion for TI bursts on HMYSOs with mass $M_*=20$~$\msun$ (see bottom panel in Fig.~\ref{fig:mdot_all_masses}).
TI outbursts in this case are too long compared with the observed short ($<5$~years) outbursts, but may be relevant for longer duration ones.

Concluding, regardless of the value of $\alpha_{\rm hot}$, $\dot{M}_{\rm dep}$, and $M_*$ used in the models, TI outbursts may only be responsible for long ($\gtrsim5$ years) outbursts on HMYSOs. It is also important to note that some HMYSOs have been observed to have multiple outbursts. All the model TI outbursts studied so far were discussed as a singular event, not connected with preceding or subsequent TI outbursts.  In the next section, we discuss TI as a possible mechanism for triggering multiple outbursts.

\section{Multiplicity of the observed bursts} \label{sec:multiplicity}

The longest of the bursts observed on HMYSO so far \citep[M17 MIR,][]{2021Chen} has a duration of $>10$ years and is still ongoing. Its duration and energy are both a few times longer than that of the other outbursts observed in HMYSOs. The timescales of this outburst are comparable to the FU Ori phenomenon in low-mass protostars. 
Furthermore, unusually for the outbursts observed on HMYSOs so far, M17 MIR has two accretion outbursts with duration of $\sim9-20$~yr each, separated by a 6 yr quiescent phase. Thus, during the 28 years of the available observations of this source, it spent most of the time in the outburst state (about $\sim80$~per~cent)  \citep{2021Chen}.  Multiple outbursts may not be unique to M17 MIR. Based on 30 years of sparse K-band photometry, a recent study by \citet{2023Fedriani} suggests that S255IR NIRS3 might have had another outburst, previously unreported, in the late 1980s, many decades before the well-known accretion outburst that was observed in 2015. This may indicate that accretion outbursts in a significant fraction of HMYSOs may have short ($\sim$ decades) repetition times.

In principle, molecular outflows triggered during the accretion outbursts might be observed as distinct outflow knots along the outflow axis. The spacing between knots can be used to determine the time interval between each individual accretion burst, assuming that each knot is associated to an accretion burst \citep[e.g.,][]{2018VorobyovElbakyanPlunkett}. It was recently shown by \citet{2024ZhouChen} that four blue-shifted CO knots along the outflow axis from M17 MIR indicate four clustered accretion outbursts occurring over the past few hundred years, each lasting for tens of years, with the intervals between these clustered accretion outbursts also being approximately tens of years.

\subsection{Thermal instability bursts with higher frequency}\label{sec:results}

\begin{figure}
\begin{centering}
\includegraphics[width=1\columnwidth]{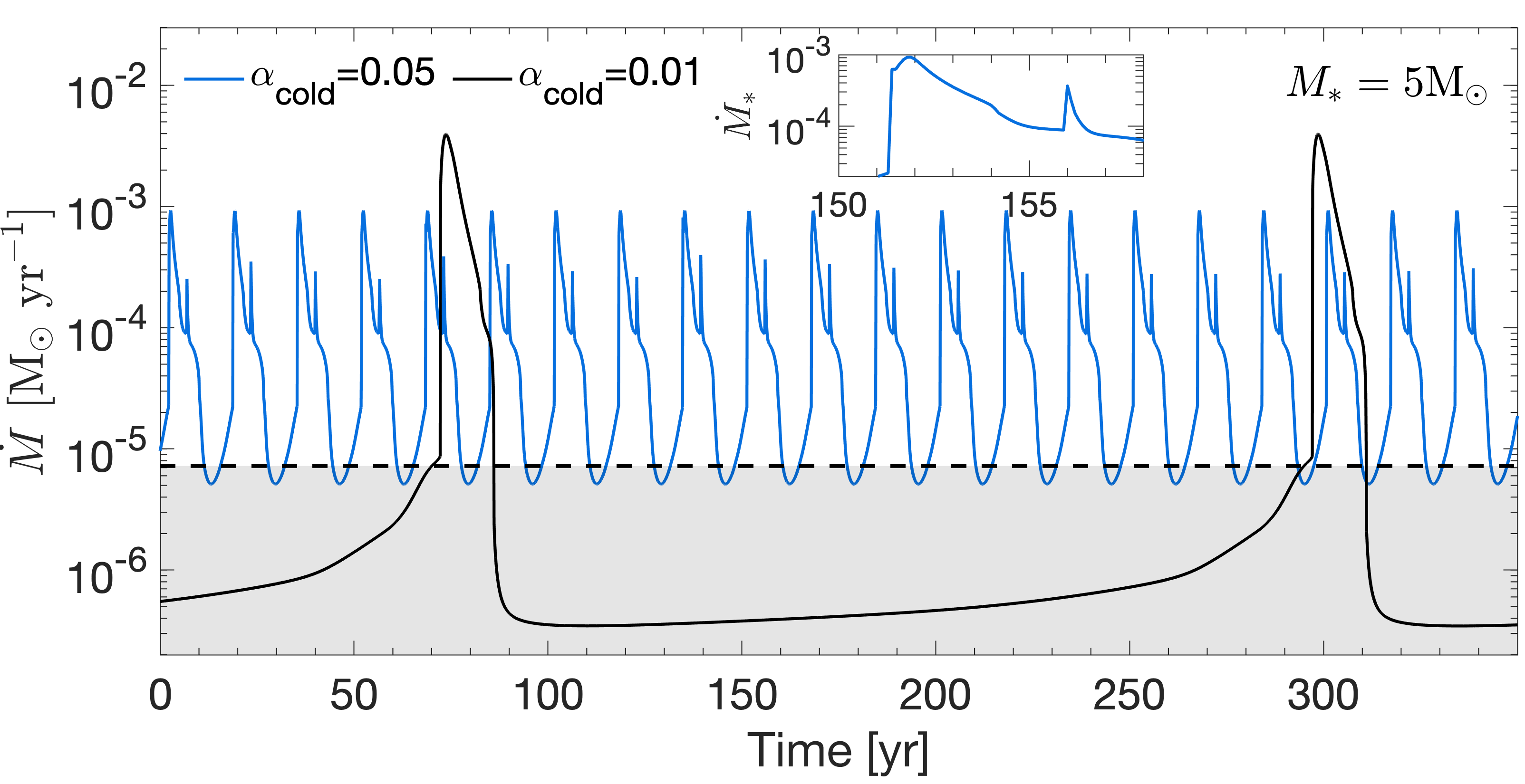}
\includegraphics[width=1\columnwidth]{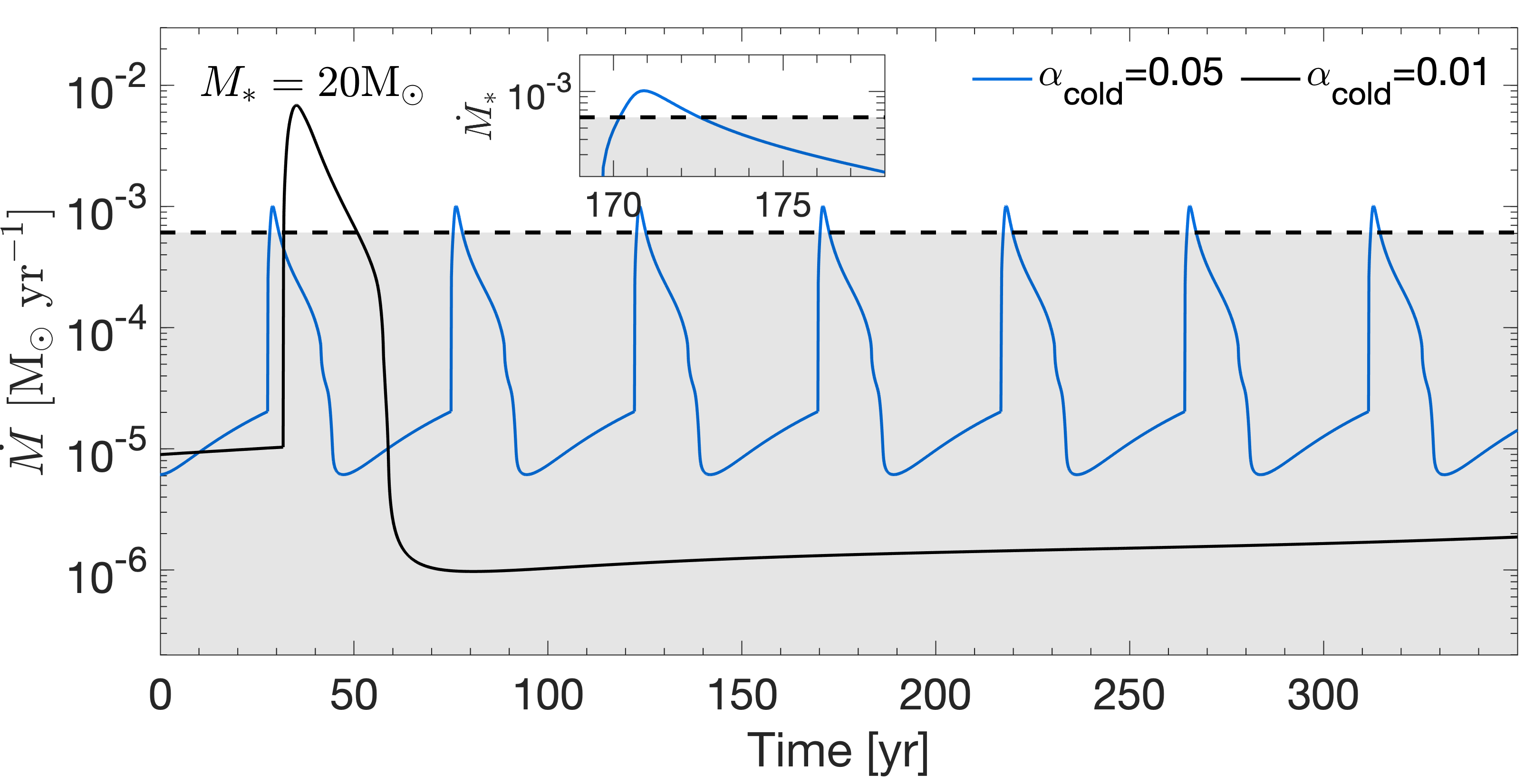}
\par\end{centering}
\caption{\label{fig:arate_alpha_cold} Accretion rate histories for the models with different $\alpha_{\rm cold}$ parameter. In all models, we use $\alpha_{\rm hot}=0.1$ and $\dot{M}_{\rm dep} = 10^{-4}$~$\msun$~yr$^{-1}$.}
\end{figure}

TI bursts in all of our models explored earlier in the paper have periodicity of more than a few 100 years, much too long to account for the  multiple bursts in M17 MIR and S255IR NIRS3. However, the repetition timescale for classical TI bursts scales with $\alpha_{\rm cold}$ as $t_{\rm rep}\propto \alpha_{\rm cold}^{-0.8}$ \citep[see Sect. 3.4 in][]{2024Nayakshin}. Therefore, here we explore if larger values of cold disc viscosity could help the TI scenario to produce rapidly repeating bursts.

In the top panel of Figure~\ref{fig:arate_alpha_cold}, we show the TI accretion bursts in HMYSO with 5~$\msun$ in two models with different $\alpha_{\rm cold}$, but with the same $\alpha_{\rm hot}=0.1$. We choose the value $\alpha_{\rm cold}=0.05$ for one of the models, since this value is higher than the conventional value of $\alpha_{\rm cold}=0.01-0.02$, but still lays close to the values of $\alpha_{\rm cold}$ obtained from the 3D MHD simulations \citep{2016Coleman, 2014Hirose}. Clearly, the model with higher $\alpha_{\rm cold}$ has more frequent TI bursts compared to the model with the classical $\alpha_{\rm cold}=0.01$ value. The periodicity of TI bursts in the higher $\alpha_{\rm cold}$ model is 17 years. This is much shorter than the periodicity of TI bursts in $\alpha_{\rm cold}=0.01$ model - 225 years, but still longer than the duration of the quiescent period between the outbursts on M17 MIR - 6 years. The higher $\alpha_{\rm cold}$ also impacts the peak accretion rate during the outburst, making it lower \citep{2001Dubus}. The peak mass accretion rates during the TI bursts in the $\alpha_{\rm cold}=0.05$ model is $10^{-3}$~$\msun$~yr$^{-1}$. Thus, even the higher $\alpha_{\rm cold}$ value shortens the quiescent period between the TI outbursts, it also makes the bursts weaker and still not consistent with the observed counterparts. In the inset panel, we show a zoom-in on one of the TI bursts in the higher $\alpha_{\rm cold}$ model. The bursts in the model has durations of a few years, but the interesting feature is the secondary burst happening during the main TI burst. This secondary burst is discussed later in Sect.~\ref{sec:reflares}. Interestingly, about a year after the outburst, a relatively weak secondary peak is observed in the S255 light curve, showing also some spectroscopic change around those epochs \citep{2020Uchiyama}.

We conduct a similar analysis of $\alpha_{\rm cold}$ impact onto the TI outbursts for our model with 20~$\msun$ mass. The results are shown in the bottom panel of Figure~\ref{fig:arate_alpha_cold}. The TI outbursts in the model with higher $\alpha_{\rm cold}$ show similar behaviour as its counterparts on 5~$\msun$ mass HMYSO discussed earlier. However, due to the higher $\dot M_{\rm acc}$, only a short period of outburst is observable in the $\alpha_{\rm cold}=0.05$ model. The duration of the observable part of the burst is about 2 years, being consistent with the observed duration on S255IR NIRS3 (2-2.5 years), but not with the ones on G323.46-0.08 (8.4 years). Moreover, the amplitude of the observable part of the model burst changes only by a factor of $\approx1.5$ in contrast to the observed change in HMYSOs by more than a factor of 5 or even an order of magnitude. 
The quiescent period between the TI bursts in $\alpha_{\rm cold}=0.05$ is about 50 years, which is about two times longer than the period of $25-30$ years between the possible bursts in S255 NIRS3. Thus, the models with higher $\alpha_{\rm cold}$ are not able to reproduce the multiple outbursts on HMYSOs with the observed properties. The burst properties in the models with various $\alpha_{\rm cold}$ are summarized in Table~\ref{tab:sum_table_alpha_cold}.

In the next section, we discuss the reflare mechanism responsible for the secondary outbursts in the model $\alpha_{\rm cold}=0.05$ with 5~$\msun$ mass and investigate the possibility of explaining the observed outburst multiplicity with the secondary outbursts.

\begin{table}[]
    \centering
    \begin{tabular}{ccccc}
    \hline 
    \hline 
    Mass & $\alpha_{\rm cold}$ & $\dot{M}_{\rm peak}$ & Burst duration & $t_{\rm rep}$ \tabularnewline
    $[M_{\odot}]$ &  & $[M_{\odot} \rm{yr}^{-1}]$ & [yr] & [yr] \tabularnewline
    \hline 

    \multirow{2}{*}{5} & 0.05 & $10^{-3}$ & 10 & 17 \tabularnewline
    & 0.01 & $3.9\times10^{-3}$ & 13.5 & 225 \tabularnewline
\tabularnewline
    \multirow{2}{*}{20} & 0.05 & $10^{-3}$ & 2 & 50 \tabularnewline
    & 0.01 & $6.6\times10^{-3}$ & 18 & $\sim$710 \tabularnewline
    \hline 
    \end{tabular}
    \caption{Burst properties in the models shown in Figure \ref{fig:arate_alpha_cold}.}
    \label{tab:sum_table_alpha_cold}
\end{table}

\subsection{Reflares}\label{sec:reflares}

\begin{figure}
\begin{centering}
\includegraphics[width=1\columnwidth]{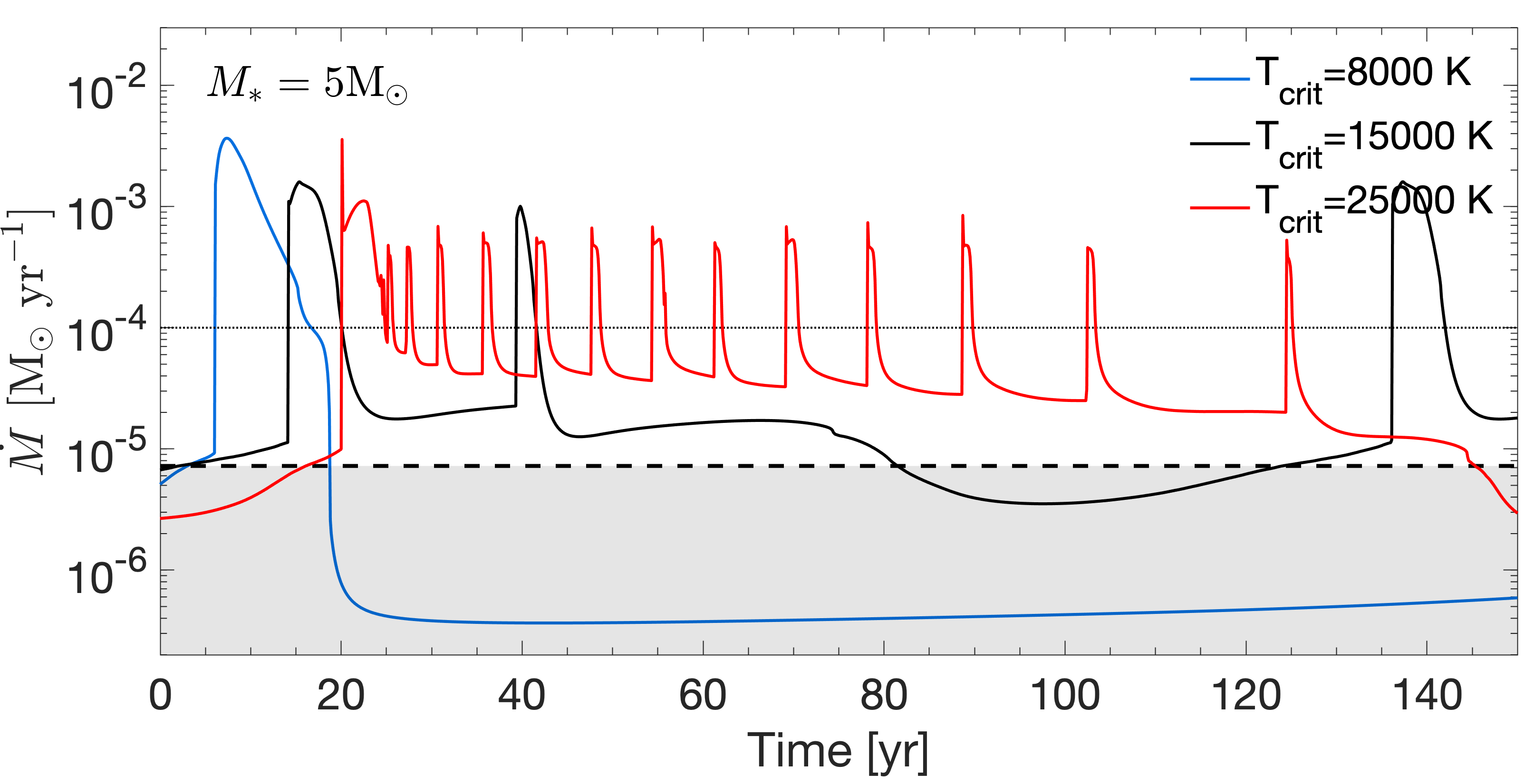}
\includegraphics[width=1\columnwidth]{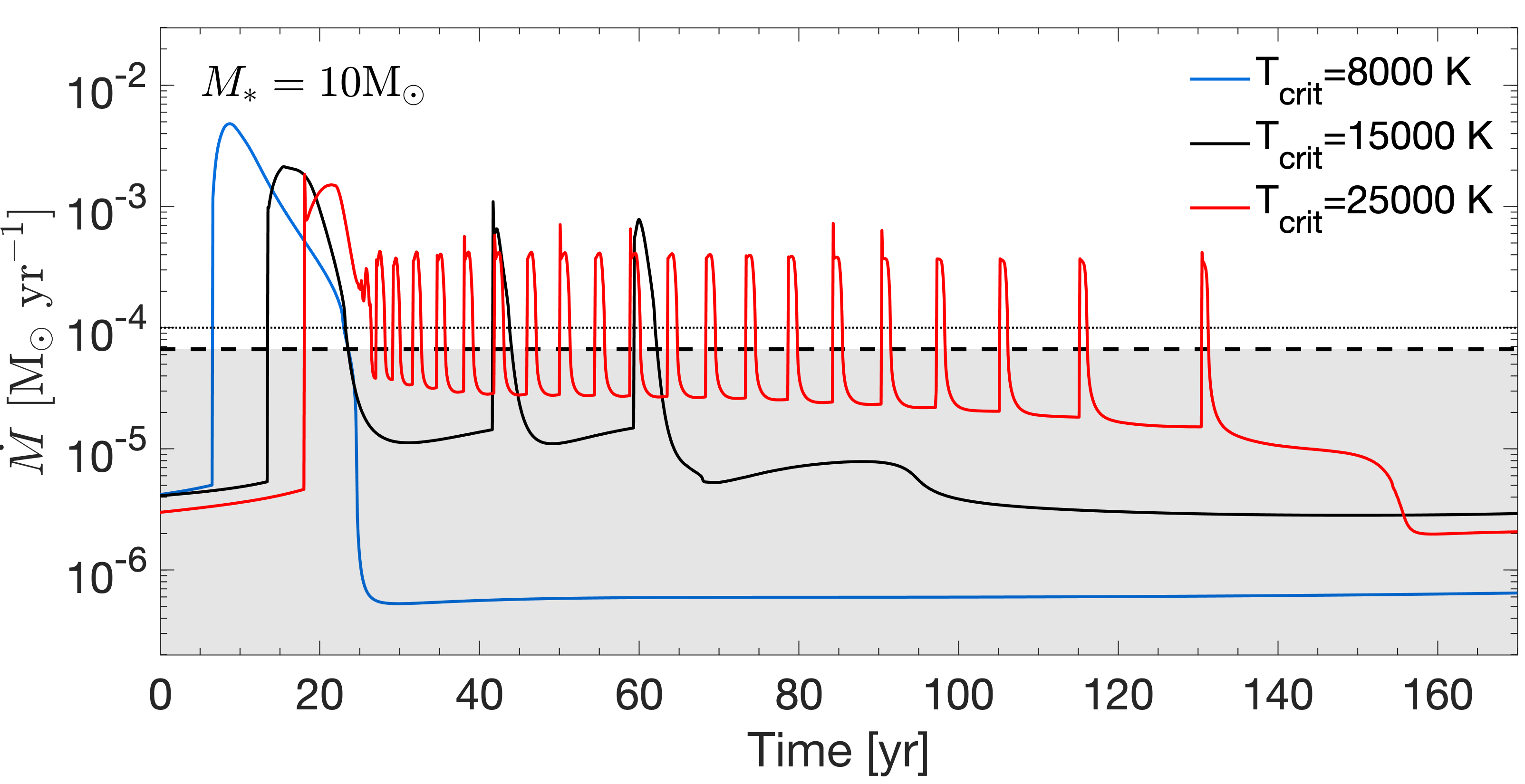}
\includegraphics[width=1\columnwidth]{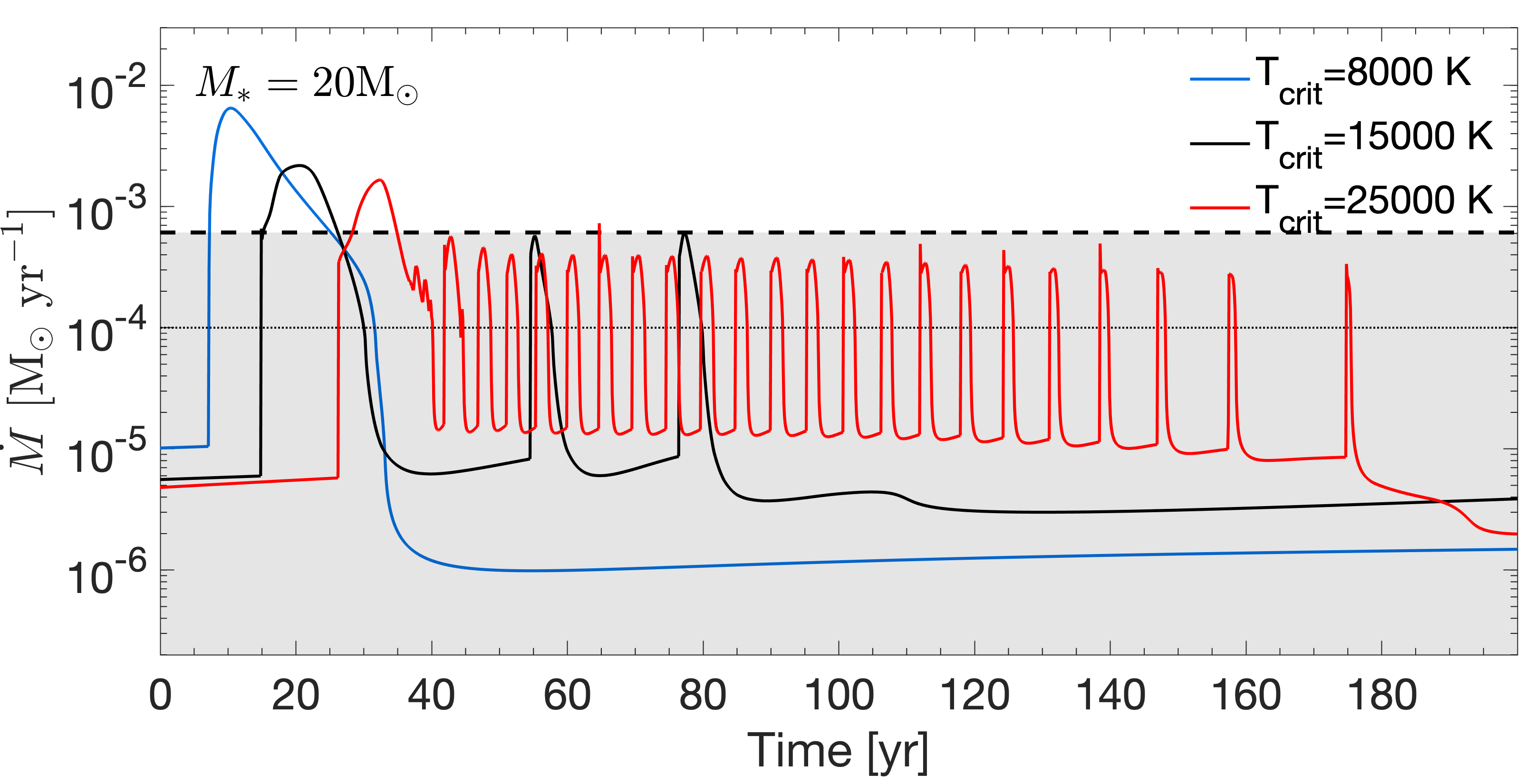}
\par\end{centering}
\caption{\label{fig:mdot_all_Tcrit} Accretion rate histories for the models with different stellar masses and with three different values of $T_{\rm crit}$ -- 8000 K (blue line), 15000 K (black line), and 25000 (red line) for each stellar mass. In all models, we use $\alpha_{\rm hot}=0.1$ and $\dot{M}_{\rm dep}=10^{-4}$~$\msun$~yr$^{-1}$. }
\end{figure}

The occurrence of multiple secondary outbursts or so-called reflares during the outburst decay is a well-known feature of TI models of discs \citep{2000Menou, 2001Dubus}. Reflares arise when the surface density behind the cooling front of the initial TI outburst achieves the critical density at which the disc enters a state of thermal instability, initiating the development of a new heating front. This front expands outward in the inside-out manner, reheating the disc until surface density falls below the critical value, allowing cooling to recommence. The reflares however do not typically occur in dwarf novae or low-mass X-ray binary transients \citep{2016Coleman}; they do appear in accretion discs around a neutron star but at very small radii \citep{2001Dubus}. Furthermore, recently \cite{2024Nayakshin} argued that reflares may have been observed in several "unusual" (that is, lasting only a few years rather than centuries) FUor outbursts on stars with mass $M_*\lesssim 1\msun$. It is therefore interesting to consider whether reflares are present in models of TI in discs around high mass stars, and whether they are observable.

The emergence of reflares and their characteristics depend on multiple factors, e.g., disc viscosity, the size of unstable region, disc irradiation by the central object, magnetic torques, and the shape of the S-curve \citep{2024Nayakshin}. For high mass stars, magnetospheric torques on the disc are likely to be negligible, because only $7\pm3\%$ of high-mass stars show large-scale surface magnetic fields \citep{2017Grunhut}.

We start from varying the $T_{\rm crit}$ parameter in our models. As shown in \citet{2024Nayakshin}, the unstable branch of S-curve shifts closer to the hot stable branch as $T_{\rm crit}$ increases, supporting the presence of reflares after the initial TI burst's decline. In the top panel of Figure~\ref{fig:mdot_all_Tcrit} we show the short period of accretion rate histories for three different models, during which a TI outburst is present. All three models have stellar mass of 5~$\msun$ and $\dot{M}_{\rm dep}=10^{-4}$~$\msun$~yr$^{-1}$, however they differ by the value of $T_{\rm crit}$. Clearly, the higher $T_{\rm crit}$ models show reflares. In the model with $T_{\rm crit}=15000$~K (black line), only one reflare is triggered, which takes place about 25 years later than the initial TI burst. This quiescent time period between the bursts is much longer than the one for M17 MIR (6~years). The durations of the initial burst and the reflare in this model are $\approx9$ and 4 years, respectively. The duration of the reflare is at least by a factor of 2 shorter than the one observed for M17 MIR ($\gtrsim9$~years). In order to decrease the quiescent period between the bursts, we recalculate the model, increasing $T_{\rm crit}$ to 25000~K (red line). Clearly, multiple reflares are present in this model with the quiescent period between the bursts varying from 3 to 22 years, which are in the range of the observed values. However, the durations of the bursts are only $2-3$~years (a factor of a few shorter than the observed ones) and the peak accretion rates during the bursts are $\approx6\times10^{-4}$~$\msun$~yr$^{-1}$ (a factor of few lower than the observed ones). This means that the reflares are unlikely to be responsible for the multiple outbursts on HMYSOs with 5~$\msun$ mass, like M17 MIR.

We carry out a similar study for the 20~$\msun$ mass HMYSO and the results are shown in the bottom panel of Figure~\ref{fig:mdot_all_Tcrit}. Similar to the 5~$\msun$ mass models shown earlier, the reflares are emerging once the value of $T_{\rm crit}$ in the model is increased. However, due to the high $\dot M_{\rm acc}$, the reflares are not observable and only the initial TI outburst is observable. Thus, the reflares are most probably not responsible for the multiple outbursts on HMYSOs with 20~$\msun$ mass.

Multiple bursts were not so far observed on HMYSOs with 10~$\msun$, however, we calculate similar models with 10~$\msun$ and show the results in the middle panel of Fig.~\ref{fig:mdot_all_Tcrit}. These results could possibly serve as a prediction for future  outburst observation on the 10~$\msun$ HMYSOs. In the model with $T_{\rm crit}=15000$~K (black line), two reflares are triggered after the initial TI burst. The reflares are separated by a quiescent time period of 18 years and have durations of about 3 years. The peak accretion rate for the first reflare is $\dot{M}_{\rm peak}\approx10^{-3}$~$\msun$~yr$^{-1}$, which is close to the observed values, while the second reflare has $\dot{M}_{\rm peak}\approx7\times10^{-4}$~$\msun$~yr$^{-1}$. Thus, the first reflare has properties close to that of observed in one of the HMYSOs with stellar mass 10~$\msun$, V723 Car. However, the burst observed in other HMYSO with stellar mass 10~$\msun$, G358.93-0.03 MM1, has a duration of a few months, which could not be reproduced with TI outbursts or the reflares in this model. In order to make the duration of the reflares shorter and try to fit the observations, we calculate another model with $T_{\rm crit}=25000$~K (red line). As in the case of 5~$\msun$ and 20~$\msun$ models, multiple reflares are triggered, which have durations of about 1 year and quiescent periods between the reflares increase with time from $\approx1.5$~years to about $\approx15$~years. However, the peak accretion rate values of the reflares are much lower than observed, being in average about $\dot{M}_{\rm peak}\approx4\times10^{-4}$~$\msun$~yr$^{-1}$. This again lead to the conclusion, that the observed short (<1 year) outbursts are unlikely to be caused by the TI or the reflares. Although, some observed outbursts could be a result not of the initial TI outbursts (which we overlook), but the subsequent reflare. The burst properties of the models with various $T_{\rm crit}$ are summarized in Table~\ref{tab:sum_table_tcrit}.

\begin{table}[]
    \centering
    \begin{tabular}{ccccc}
    \hline 
    \hline 
    Mass & $T_{\rm crit}$ & $\dot{M}_{\rm peak}$ & Burst duration & $t_{\rm rep}$ \tabularnewline
    $[M_{\odot}]$ & [K] & $[M_{\odot} \rm{yr}^{-1}]$ & [yr] & [yr] \tabularnewline
    \hline 

    \multirow{3}{*}{5} & 8000 & $3.9\times10^{-3}$ & 13.5 & $\sim$300 \tabularnewline
    & 15000 & $10^{-3}$ & 9.4 & 25 \tabularnewline
    & 25000 & $6\times10^{-4}$ & 2--3 & 3--22 \tabularnewline
\tabularnewline
    \multirow{3}{*}{10} & 8000 & $5\times10^{-3}$ & 17.8 & $\sim$450 \tabularnewline
    & 15000 & $10^{-3}$ & 3 & 18 \tabularnewline
    & 25000 & $4\times10^{-4}$ & 1 & 1.5--15 \tabularnewline
\tabularnewline
    20 & \multicolumn{4}{c}{Reflares are not observable} \tabularnewline

    \hline 
    \end{tabular}
    \caption{Burst properties in the models shown in Figure \ref{fig:mdot_all_Tcrit}.} 
    \label{tab:sum_table_tcrit}
\end{table}

\section{Discussion}\label{sec:discussion}

\subsection{Comparison with the photometric data of S255IR NIRS3} \label{sec:photometric}

\begin{figure}
\begin{centering}
\includegraphics[width=1\columnwidth]{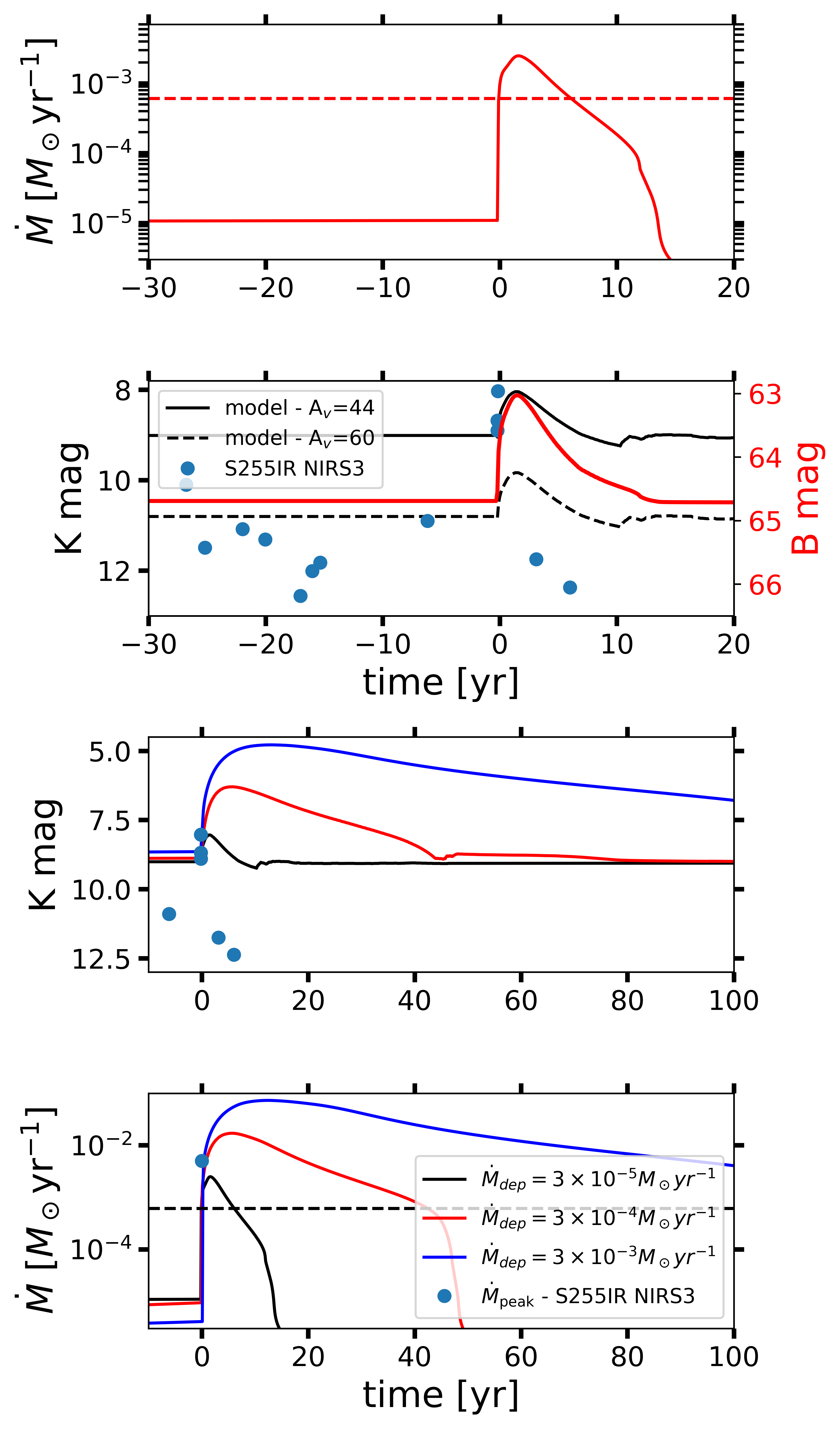}
\par\end{centering}
\caption{\label{fig:compare_obs} \textbf{Top panel:} Mass accretion rate onto the star of mass $M_* = 20$~$\msun$ during a TI outburst. The horizontal dashed line shows the $\dot M_{\rm th}$ value calculated with Eq.~(\ref{eq:mdot_obs}). \textbf{Second panel:} Model K band magnitudes during the TI outbursts (solid black line) and the observed K band photometry for S255IR NIRS3 source from \citet{2023Fedriani} (blue dots). The K band magnitudes for the model with the optical extinction $A_v=60$ is shown with the dashed black line. The red line corresponds to the model B band magnitudes. The time $t=0$ corresponds to the beginning of the model TI outburst and to the beginning of the observed outburst on S255IR NIRS3 registered in June 2015. 
\textbf{Third panel:} Model K band magnitudes for the model presented in the top two panels of the figure (black line) and for two other models with the higher $\dot{M}_{\rm dep}$. The blue dots represent the observational data also shown in the second panel. \textbf{Bottom panel:} Mass accretion rate histories during the TI bursts in the models shown in the third panel with the corresponding colour. The horizontal dashed line marks the $\dot M_{\rm th}$ value. The time $t=0$ is similar to the one in the top two panels. The blue dot marks the observed peak mass-accretion rate value on S255IR NIRS3 during the outburst. }
\end{figure}

One of the important feature in the S255 NIRS3 burst is the increase of luminosity in K band ($\lambda=2.2\mu$m) by $>2$~mag \citep{2017Caratti}.
In order to compare our model results with the observational data, we generate integrated black body spectra of the disc using our time-dependent models and determine the observed magnitudes of the disc within our computational domain, excluding the star and the envelope. For the source distance, we assume 1.78 kpc \citep{2016Burns} and for the optical extinction ($A_V$) a value of 44 mag \citep{2017Caratti}.

In the top panel of Figure~\ref{fig:compare_obs}, we show the mass accretion rate during one of the TI outbursts on HMYSO with the mass of 20~$\msun$. The horizontal dashed line shows the $\dot M_{\rm th}$ value. The black solid line in the second panel shows the model magnitude in the photometric K band ($\lambda=2.2\mu$m) during the TI outburst. The blue dots in the second panel show the K band photometry data for S255IR NIRS3 source obtained from \citet{2023Fedriani}. The time instance $t=0$ corresponds to the beginning of model TI outburst and the confirmed outburst registered in June 2015 on S255IR NIRS3 source. Clearly, in the model, the increase in K band due to the outburst is $\sim1$~magnitude, while the observed K band for S255IR NIRS3 increased for a few magnitudes during the outburst. Meanwhile, the increase in the mass accretion rate during the model TI outburst is consistent with the change of mass accretion rate for S255IR NIRS3 during the outburst, estimated from the observations. Thus, even if the model peak accretion rates during the burst and the burst duration are similar to that of observed, a strong inconsistency can still be present between the model and observational luminosities in some photometric bands. 

We note that the model pre-burst K band magnitudes are much higher than the ones observed if we use $A_v=44$ mag. However, the estimated error range for the observed value of $A_v$ is quite large, being $\pm16$ mag\citep{2017Caratti}. If we use the highest value in this error range, $A_v=60$ mag, then the pre-burst K band magnitudes in our model fits very well with the observations. K band magnitudes for the model with higher $A_v$ are shown with the dashed line in the second panel of Figure~\ref{fig:compare_obs}. The increase in the K band during the outburst is again only about a magnitude, which is much lower than that of observed. This result emphasizes the importance of not only comparing model mass accretion rates during the bursts and their durations, but also conducting photometric analysis to comprehensively assess the agreement between model predictions and observational data.

Some of the observed variability (a few magnitudes) in K band may also be due to the variability of $A_V$ during the burst. Extinction variation (pre- or post-burst) is observed for some FUOrs \citep{2024Guo}, and often attributed to dust grain condensation when the wind collides with the envelope \citep{2023Siwak} or to dust ejection during the bursts. This observed change in extinction occurs on short timescales of years.

The reason for the low variability in model K band is that the effective temperatures corresponding to this wavelength ($\sim1300$~K) are present in the disc at the radial distance of $\sim1-2$~au. At this radius, the influence of the TI outburst is much weaker, compared to the sub-au distances in the disc. For the comparison, we present in the second panel of Figure~\ref{fig:compare_obs} the model B band magnitudes (red line), which correspond to the inner and hotter regions of the disc ($\sim6600$~K). Clearly, the increase in B band is stronger, reaching about 2 magnitudes.

In order to show which TI bursts will ensure the change in the K band for more than 2 magnitudes in a HMYSO of 20 $\msun$, we vary the value of $\dot{M}_{\rm dep}$ in our model, while fixing the rest of the model parameters. The results for these models are shown in the bottom two panels of Figure~\ref{fig:compare_obs}. The black line in the bottom two panels correspond to the model with $\dot{M}_{\rm dep}=3\times10^{-5}$~$\msun$~yr$^{-1}$ shown in the top two panels. The two other models with $\dot{M}_{\rm dep}=3\times10^{-4}$~$\msun$~yr$^{-1}$ and $\dot{M}_{\rm dep}=3\times10^{-3}$~$\msun$~yr$^{-1}$ are presented in the bottom two panels of the figure with red and blue lines, respectively. Clearly, to have an increase for $>2$ magnitudes in K band during a TI outburst, one should have $\dot{M}_{\rm dep}>$ a few $\times10^{-4}$~$\msun$~yr$^{-1}$. The latter values are in agreement with the observed values for the outbursting HMYSOs \citep{2013Beuther, 2021Moscadelli}. However, as it can be seen from the bottom panel of Figure~\ref{fig:compare_obs}, the peak accretion rates during such a TI outburst will have values $\dot{M}_{\rm peak}\gtrsim10^{-2}$~$\msun$~yr$^{-1}$, which is much larger than the observed values of peak accretion rates in S255IR NIRS3 source during its outburst event (blue dot). Moreover, the burst duration in the models with higher $\dot{M}_{\rm dep}$ increases to a few tens (red line) and more than a hundred years (blue line), which are much longer than the observed duration of the bursts in S255IR NIRS3. Another burst feature that becomes strongly inconsistent with the observations for the higher $\dot{M}_{\rm dep}$ models is the rise time of the burst. The rise time also increases and becomes a few years (red line) and more than a decade (blue line), which is much longer than the observed value of 0.4 years for S255IR NIRS3. Thus, the observed strong variability ($>$2 magnitudes) in K band during a short period of time ($\sim2$~yrs) in S255IR NIRS3 source most likely is not a result of TI burst in the inner disc.

\subsection{Possible mechanisms responsible for bursts on HMYSOs} \label{sec:alter_mech}
Our analysis of TI models in the context of HMYSO outbursts reveals their potential to explain some key characteristics such as burst duration and magnitude. However, several critical discrepancies emerge when comparing model predictions with observed phenomena.

One notable limitation of current TI models is their inability to fully account for the multiplicity of bursts observed in some HMYSOs. While the simulations accurately reproduce individual outbursts in terms of duration and magnitude in some systems, they fail to capture the sequential or recurrent nature of burst occurrences observed in other systems. This discrepancy suggests that additional factors or mechanisms may be at play in regulating the burst frequency and timing in HMYSOs.

One possible avenue for addressing this inconsistency is the consideration of external influences, such as gravitational interactions or disc instabilities. In addition to TI, gravitational instability (GI) and disc fragmentation have been proposed as alternative mechanisms for triggering the observed outbursts in HMYSOs \citep{2017MeyerVorobyov, 2020OlivaKuiper, 2021ElbakyanNayakshin, 2023ElbakyanNayakshin}. In this scenario, the disc undergoes fragmentation into clumps, which can then accrete onto the central protostar, triggering episodic bursts of accretion and mass ejection. While thermal instability remains a candidate for explaining HMYSO outbursts, the roles of gravitational instability and disc fragmentation warrant further investigation. Future studies incorporating these mechanisms into comprehensive numerical simulations could provide valuable insights into the relative contributions of each process to the observed burst phenomena. By exploring a diverse range of physical conditions and initial parameters,  we can assess the viability of gravitational instability and disc fragmentation as complementary or alternative mechanisms for driving episodic accretion in HMYSOs.

Furthermore, the observed diversity in burst characteristics among different HMYSOs underscores the need for a more nuanced approach to model these phenomena. Future efforts should aim to develop more sophisticated TI models that can account for the full range of observed behaviours, including both singular and multiple burst events. Incorporating observational constraints from a broader sample of HMYSOs will be crucial for refining and validating these models, ultimately advancing our understanding of the underlying processes driving star formation and evolution.

\subsection{Parameter space study}
Additionally, it is important to note that our TI models contain several free parameters that are adjusted to match observed burst characteristics. While we have explored a range of parameter values by varying them, it is impractical to comprehensively study all possible combinations of parameters due to the vastness of the parameter space. However, it is worth highlighting that the diverse combinations of free parameters in the models can yield new and interesting results, potentially revealing novel insights into HMYSO outburst phenomena.

The exploration of different parameter sets may uncover previously unexplored regimes of behaviour, shedding light on the underlying physical processes driving burst initiation and evolution. While our current study focuses on a subset of parameter values that produce reasonable agreement with some observed burst characteristics, a comprehensive exploration of the parameter space remains a valuable avenue for future research. Such an endeavour would involve systematic investigations of a broader range of parameter values, potentially uncovering new regimes of behaviour and refining our understanding of HMYSO outburst mechanisms. By leaving this comprehensive parameter space study for future work, we open the door to further advancements in our modelling efforts and a deeper understanding of the complex interplay between physical processes governing HMYSO evolution.

\subsection{1D vs multi-D models}
We acknowledge the limitations of our 1D modelling of HMYSO accretion disks. While 1D models provide valuable insights into the basic physical processes governing disc evolution and outburst phenomena, they inherently oversimplify the complex interplay of dynamics occurring within the disc. The transition to 2D or 3D modelling represents a crucial step toward capturing the intricate dynamics of HMYSO accretion disks, allowing for a more realistic treatment of phenomena such as disc fragmentation, gravitational instabilities, and the formation of protostellar companions. By incorporating azimuthal variations in disc structure and evolution, multidimensional models offer a more accurate representation of the processes driving episodic accretion in HMYSOs. Future studies employing multidimensional hydrodynamic simulations are essential for elucidating the role of disc morphology, turbulence, and non-axisymmetric structures in shaping the observed burst behaviour of HMYSOs. First steps in this direction are already done by \citet{2024Jordan} with the new FargoCPT code \citep{2024Rometsch}. The authors simulate complete outburst cycles using the thermal tidal instability model in non-isothermal, viscous accretion disks in cataclysmic variable systems.

\section{Conclusions}\label{sec:conclusion}

In this paper, using the 1D hydrodynamical code DEO, we studied the phenomenon of thermal instability (TI) outbursts on high-mass young stellar objects (HMYSOs) and their implications for the early stages of massive star formation.  
This modelling allowed us to quantify TI outburst properties and their dependence on model parameters, such as  stellar mass and the viscosity prescription. Our main conclusions are: 
\begin{itemize}

\item Our simulations reproduce the main characteristics of the longer bursts observed in HMYSOs reasonably well. However, they fail to reproduce the short-duration (less than a few years) outbursts with the short rise times (a few weeks or months) observed in several HMYSOs. This suggests another mechanism, such as disc fragmentation, as a potential alternative in generating these shorter outbursts (cf. Sect.~\ref{sec:alter_mech}).

\item Despite exploring a wide parameter space in $\alpha_{\rm hot}$, $\dot{M}_{\rm dep}$, and $M_*$, our TI models also struggle to reproduce multiple outbursts observed in a few sources (Sect.~\ref{sec:obser} \& \ref{sec:multiplicity}). Notably, the recurrent outbursts in observed systems like M17 MIR and possibly S255 NIRS3, characterized by sequential bursts separated by relatively short quiescent phases, pose a challenge to the current TI models. More specifically, the repetition timescale of model outbursts is too long to be humanly observed to repeat (cf Table~\ref{tab:sum_table}.) This discrepancy signals the need for a refined or alternative mechanism to comprehensively elucidate the observed multiplicity and frequency of bursts in HMYSOs.

\item SED modelling of TI outbursts  (Sect.~\ref{sec:photometric}) shows that TI outburst model luminosity in various photometric bands can strongly differ from the observed values, despite the model displaying $\dot M_*$ similar to that inferred observationally. This indicates that the model flux as a function of radius in the disc is different from that of the system modelled, and is therefore an important tool that modellers should not neglect to use.

\item While our model TI outbursts do not offer a comprehensive explanation for the whole range of the observed episodic accretion variability in HMYSOs, they do show TI to be an important part of disc physics of these objects. TI sculpts the structure of the inner few au of HMYSO discs and  cannot be simply discounted as "unwanted as not quite successful". Any other accretion variability agent, such as e.g., massive migrating clumps, must contend with the effects that TI has on the inner disc. This should not be viewed as a modelling nuisance but be instead embraced as an additional tool to probe potential scenarios for HMYSO accretion outbursts.

\end{itemize}

Our study demonstrates that TI accretion outbursts are a common occurrence in high-mass YSOs, during their early stages of formation. These bursts of accretion provide important insights into the mass accretion process, the evolution of circumstellar disks, and the formation of multiple star systems. Further investigations combining observational studies with theoretical modelling will contribute to a more comprehensive understanding of the complex processes involved in the birth of massive stars.

\section{Acknowledgement}
We thank the anonymous referee for their thoughtful and insightful feedback, which has greatly improved the quality of this paper. V.E. and S.N. acknowledge the funding from the UK Science and Technologies Facilities Council, grant No. ST/S000453/1. This work made use of the DiRAC Data Intensive service at Leicester, operated by the University of Leicester IT Services, which forms part of the STFC DiRAC HPC Facility (www.dirac.ac.uk). C.G. acknowledges support from PRIN-MUR 2022 20228JPA3A “The path to star and planet formation in the JWST era (PATH)” funded by NextGeneration EU and by INAF-GoG 2022 “NIR-dark Accretion Outbursts in Massive Young stellar objects (NAOMY)” and Large Grant INAF 2022 “YSOs Outflows, Disks and Accretion: towards a global framework for the evolution of planet forming systems (YODA)”. R.K. acknowledges financial support via the Heisenberg Research Grant funded by the Deutsche Forschungsgemeinschaft (DFG, German Research Foundation) under grant no.~KU 2849/9, project no.~445783058.


\bibliographystyle{aa}
\bibliography{refs.bib}

\label{lastpage}
\end{document}